\documentclass[lettersize,journal]{IEEEtran}
\usepackage{amsmath,amsfonts}
\usepackage{array}
\usepackage[caption=false,font=normalsize,labelfont=sf,textfont=sf]{subfig}
\usepackage{textcomp}
\usepackage{stfloats}
\usepackage{url}
\usepackage{verbatim}
\usepackage{graphicx}
\usepackage{cite}
\DeclareMathOperator*{\argmin}{arg\,min}
\usepackage{algorithm}
\usepackage{algorithmicx}
\usepackage{algpseudocode}
\algnewcommand{\LineComment}[1]{\State \(\triangleright\) #1}

\usepackage{booktabs}
\usepackage{multirow}
\usepackage[T1]{fontenc}

\usepackage{calc} % Table makebox
\usepackage[table,xcdraw]{xcolor}

\usepackage[pdfstartview=XYZ,
bookmarks=true,
colorlinks=true,
linkcolor=blue,
urlcolor=blue,
citecolor=blue,
bookmarks=true,
linktocpage=true, % makes the page number as hyperlink in table of content
hyperindex=true
]{hyperref}

\usepackage{orcidlink}

\begin{document}

\title{Hypnopaedia-Aware Machine Unlearning via Psychometrics of Artificial Mental Imagery}

\author{Ching-Chun Chang\orcidlink{0000-0001-7723-4591}, Kai Gao\orcidlink{0000-0002-7505-9037}, Shuying Xu\orcidlink{0000-0003-3082-8127}, Anastasia Kordoni\orcidlink{0000-0001-9823-2341}, Christopher Leckie\orcidlink{0000-0002-4388-0517} and Isao Echizen\orcidlink{0000-0003-4908-1860} % <-this % stops a space

\thanks{C.-C. Chang is with National Institute of Informatics, Chiyoda, Tokyo, Japan (email: {ccchang@nii.ac.jp}).}% <-this % stops a space
\thanks{K. Gao and S. Xu are with the Department of Information Engineering and Computer Science, Feng Chia University, Taichung, Taiwan (email: {kaigao.phd@gmail.com} and {shuyin.xu.phd@gmail.com}).}% <-this % stops a space
\thanks{A. Kordoni is with the Department of Psychology, Lancaster University, Lancaster, UK (email: {a.kordoni@lancaster.ac.uk}).}% <-this % stops a space
\thanks{C. Leckie is with the Department of Computing and Information Systems, University of Melbourne, Melbourne, VIC, Australia (email: {caleckie@unimelb.edu.au}).}% <-this % stops a space
\thanks{I. Echizen is with the Information and Society Research Division, National Institute of Informatics; the Department of Informatics, Graduate University for Advanced Studies; and the Department of Information and Communication Engineering, University of Tokyo, Chiyoda, Tokyo, Japan (email: {iechizen@nii.ac.jp}).}% <-this % stops a space
}

% The paper headers
\markboth{}%
{}

\maketitle

\begin{abstract}
Neural backdoors represent insidious cybersecurity loopholes that render learning machinery vulnerable to unauthorised manipulations, potentially enabling the weaponisation of artificial intelligence with catastrophic consequences. A backdoor attack involves the clandestine infiltration of a trigger during the learning process, metaphorically analogous to hypnopaedia, where ideas are implanted into a subject's subconscious mind under the state of hypnosis or unconsciousness. When activated by a sensory stimulus, the trigger evokes a conditioned reflex that directs a machine to mount a predetermined response. In this study, we propose a cybernetic framework for constant surveillance of backdoor threats, driven by the dynamic nature of untrustworthy data sources. We develop a self-aware unlearning mechanism to autonomously detach a machine's behaviour from the backdoor trigger. Through reverse engineering and statistical inference, we detect deceptive patterns and estimate the likelihood of backdoor infection. We employ model inversion to elicit artificial mental imagery, using stochastic processes to disrupt optimisation pathways and avoid convergent but potentially flawed patterns. This is followed by hypothesis analysis, which estimates the likelihood of each potentially malicious pattern as the true trigger and infers the probability of infection. The primary objective of this study is to maintain a stable state of equilibrium between knowledge fidelity and backdoor vulnerability.
\end{abstract}

%\newpage
\section{Introduction}
\IEEEPARstart{C}{ybersecurity} stands at the frontline of trustworthy artificial intelligence by addressing evolving threats and preventing malicious actions that could undermine the safety and trust in computational intelligence. Backdoors (or Trojan horses) represent concealed entry points that allow attackers to manipulate the behaviour of a machine and weaponise artificial intelligence, raising serious cybersecurity concerns~\cite{9802938}. A backdoor attack functions by infiltrating a hidden trigger into a machine during its learning phase, which, when activated, causes it to produce predetermined and often harmful responses. It forms a \emph{conditioned reflex}, an automatic and conditioned response paired with a specific stimulus~\cite{pavlov1927conditioned}.

The implications of backdoors are wide-ranging. In social computing, a backdoor could subvert ethical filters and content moderation, instructing generative artificial intelligence to create and disseminate misinformation. In autonomous vehicles, it could cause misinterpretation of traffic signals, leading to potentially catastrophic accidents. In biometric recognition, it could allow unauthorised access that bypasses security protocols. In the financial industry, fraud detection systems could be compromised, enabling fraudulent transactions under specific conditions. In the healthcare sector, medical diagnostic systems could be manipulated to deliver incorrect diagnoses and treatments. These potential consequences underscore the urgent need for robust countermeasures to prevent, detect, and mitigate the risks and threats posed by backdoors.

The dynamic and often uncontrollable nature of data sources further complicates this challenge. This is exacerbated in \emph{federated learning} (or collaborative learning) due to the presence of compromised nodes~\cite{pmlr-v54-mcmahan17a}. Federated learning enables the decentralisation of data sources, offering benefits such as promoting large-scale collaboration, preserving privacy, reducing data breach risks, improving data utilisation efficiency, and preventing monopolistic control over data. However, it also comes with risks. Malicious local participants can inject harmful data and false computations (which are not centrally verifiable), potentially introducing backdoors when aggregated into a global model. Furthermore, systems featuring \emph{lifelong learning} to continuously and incrementally adapt to new data over time may face similar challenges due to dynamic environments that involve crowdsourced data labelling and open data repositories~\cite{schmidhuber:1987:srl, Thrun:1998aa, Lake:2017aa, pmlr-v97-finn19a}. To manage these risks, developing a feedback control mechanism that continuously monitors the presence of backdoors is essential to maintaining system integrity and reliability.

\begin{figure}[!t]
\centering
\includegraphics[width=0.9\linewidth]{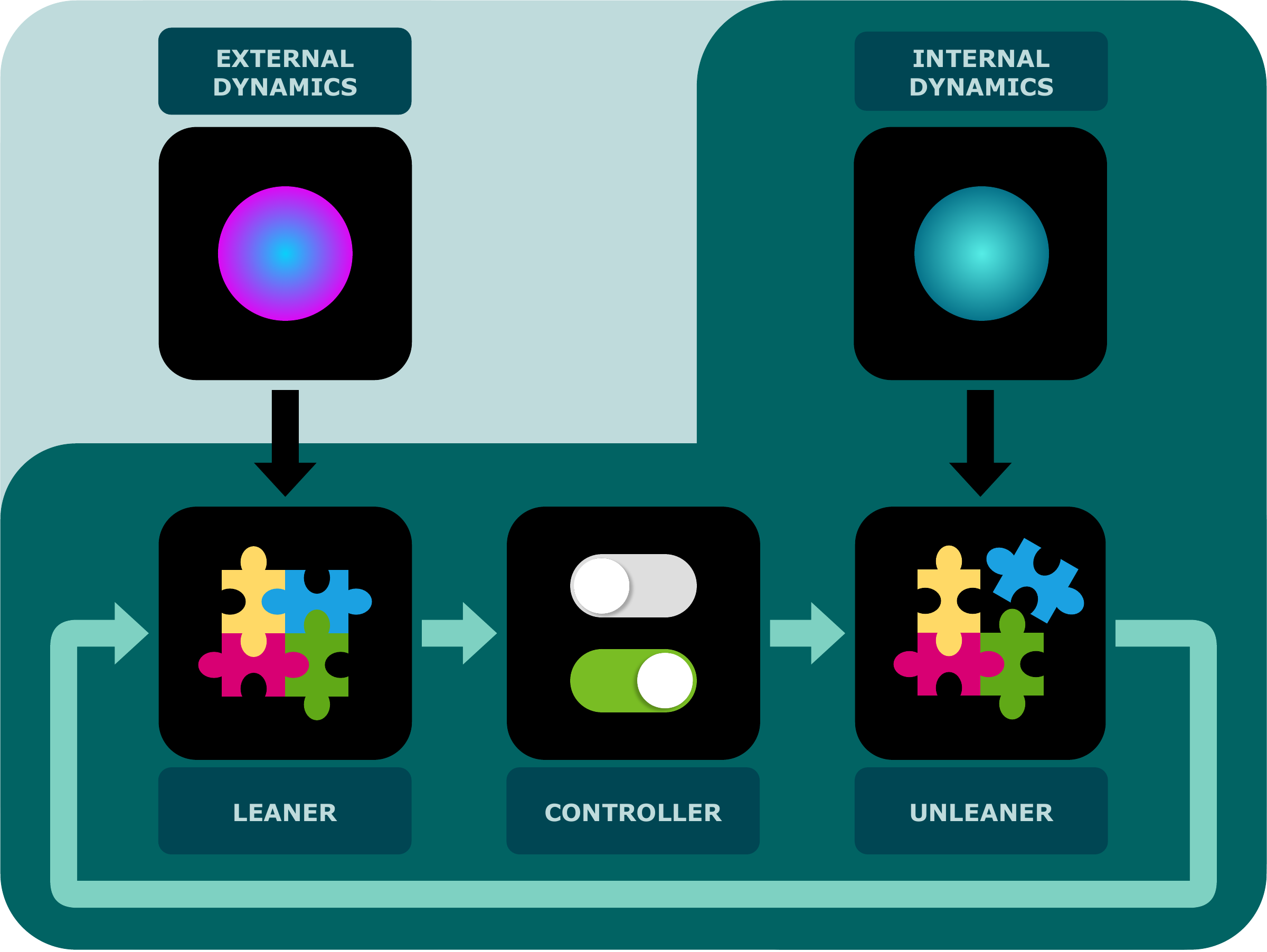}
\caption{Cybernetic framework that consists of learner, controller and unlearner for backdoor awareness.}
\label{fig:cybernetics}
\end{figure}

\begin{figure}[!t]
\centering
\includegraphics[width=0.9\linewidth]{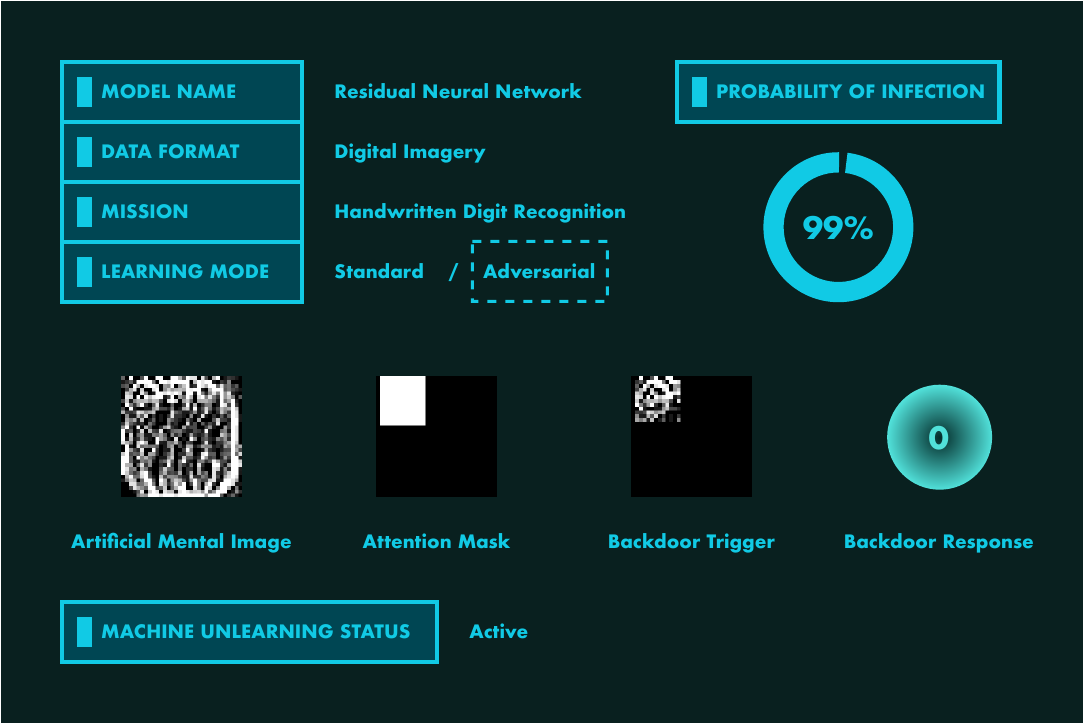}
\caption{Psychometric profile that shows probability of infection, backdoor trigger, backdoor response and auxiliary forensic information.}
\label{fig:ui}
\end{figure}

In this study, we propose a cybernetic framework for mitigating the impact of backdoors to neural machines in a continual learning scenario, as illustrated in Figure~\ref{fig:cybernetics}. It consists of a leaner which updates the machine with untrustworthy external data sources under the risks of data poisoning, a controller which steers the machine towards the decision of whether or not unlearn to unlearn, and an unlearner which updates the machine with trustworthy internal data sources and auxiliary information about the backdoor. It begins by performing \emph{model inversion} to elicit artificial mental images. A multi-scale gradient-descent optimisation algorithm is employed to synthesise artificial mental images in a coarse-to-fine manner. Next, \emph{hypothesis analysis} is conducted to identify the most likely hypothetical trigger pattern extracted from the artificial mental images using maximum likelihood estimation with outlier exclusion and infer the probability of infection using Bayesian inference. This involves scanning through all potential regions to estimate the probability of each regional pattern being the trigger, based on the machine's response to a small collection of samples. The decision to unlearn or remain intact is then made according to the psychometric profile, codenamed \emph{Psycho-Pass}, as illustrated in Figure~\ref{fig:ui}. If \emph{machine unlearning} is activated, a collection of unlearning samples is used for disassociating the hypothetical trigger and its corresponding behaviour. However, side effects lurk due to internal dynamics such as the propagation of uncertainty and stochastic biases in the data and analysis process, potentially deteriorating the performance of the machine. The research objective is to balance the dynamics between a learner agent and an unlearner agent, preserving the \emph{fidelity} of the machine while minimising its \emph{vulnerability} to backdoor attacks.

\section{Preliminaries}
In this section, we lay the foundation for understanding the landscape of backdoor attacks and defences. We begin by introducing a taxonomy that systematically categorises the diverse characteristics of backdoor attacks. Following this, we delve into both proactive and reactive defence paradigms, outlining strategies to prevent, detect and mitigate these insidious threats. To ensure clarity and relevance, we then delineate the scope of our research, specifying the attack and defence scenarios under investigation. Furthermore, we briefly review solutions for reverse engineering backdoor triggers, which serve as essential benchmarks for comparative study.

\subsection{Backdoor Attacks}
A backdoor is a deliberate vulnerability or loophole inserted into a neural network model that allows an attacker to manipulate its behaviour and compromise its functionality. This manipulation typically occurs by adding specific patterns or triggers to the input data, which the model then incorrectly identifies or responds to. In a nutshell, an attacker with access to the model's learning data or learning process injects a specific pattern or trigger into the data. This pattern could be innocuous or subtle, making it hard to detect during normal operation. Once the model is deployed and in use, the attacker can activate the backdoor by feeding in input data that contains the trigger pattern. When the model encounters this trigger, it behaves in a specific, predetermined way, often giving incorrect or malicious outputs. The consequences can vary depending on the context. In a security application, a backdoor might allow an attacker to bypass authentication systems or gain unauthorised access. In a financial application, it could manipulate predictions to favour certain outcomes, leading to fraud or financial losses.

\subsubsection*{Backdoor Taxonomy}
Understanding backdoor attacks involves several key concepts that shed light on their nature and impact. Space refers to where triggers are applied: either samples in cyberspace (digital environment), or samples in physical space (real-world environment)~\cite{9577800}. Causality defines the mappings between inputs and outputs, either as all-to-one, where multiple samples lead to a single targeted prediction, or all-to-all, where different samples may be linked to different manipulated predictions~\cite{NEURIPS2020_234e6913}. Genericity distinguishes whether triggers are uniform across different samples or specific to individual instances~\cite{9711191, 9797338, 9870671}. Optimality reflects whether triggers are arbitrary handcrafted patterns or optimised for maximum effectiveness of backdoor attacks~\cite{Trojannn, geiping2021witches, NEURIPS2022_79eec295}. Semanticity describes the relationship between triggers and the semantic content of samples, whether triggers are independent of or integrated seamlessly into samples~\cite{pmlr-v108-bagdasaryan20a}. Visibility concerns whether triggers are perceptible or designed to avoid visible distortions to samples~\cite{10.1145/3374664.3375751, 9186317, 10.1007/978-3-031-19778-9_23}. In summary, backdoor triggers can be characterised by the following taxonomic descriptions.
\begin{itemize}
	\item \textit{Space}: Triggers are applied to samples in cyberspace or physical space.
	\item \textit{Causality}: Triggers cause all-to-one or all-to-all mappings between inputs and outputs.
	\item \textit{Genericity}: Triggers are generic (same) or specific (different) for each sample.
	\item \textit{Optimality}: Triggers are arbitrary handcrafted patterns or optimised for successful attack.
	\item \textit{Semanticity}: Triggers are semantically dependent or independent parts of samples.
	\item \textit{Visibility}: Triggers are visible or invisible to human perceptual systems.
\end{itemize}

\begin{figure}[!t]
\centering
\includegraphics[width=0.9\linewidth]{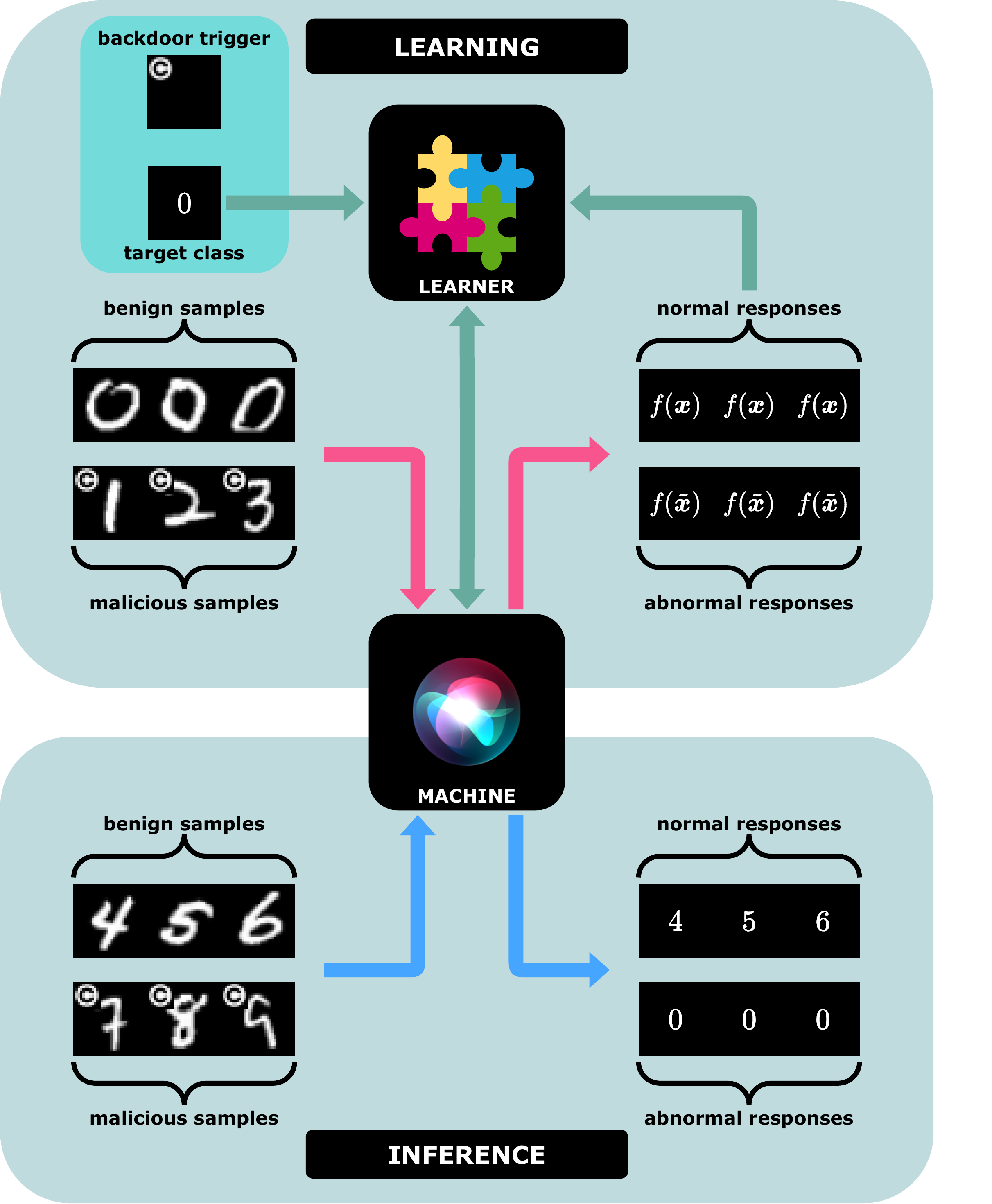}
\caption{Illustration of backdoor attack through implanting triggers into samples during the learning phase.}
\label{fig:attack}
\end{figure}

\subsection{Backdoor Defences}
Defending against backdoor attacks is crucial to ensure the integrity and security of machine learning systems. A variety of defence mechanisms have been developed to counteract backdoor attacks, which can be broadly categorised into \textit{proactive} (or learning-time) and \textit{reactive} (or inference-time) paradigms. As the names suggest, the proactive paradigm focuses on securing the learning data and process in the pre-deployment phase, whereas the reactive paradigm aims to offset the impact of backdoors in the post-deployment phase.

\subsubsection*{Proactive Paradigm}
Proactive defence is designed to prevent the insertion of backdoors or mitigate the impact of backdoors during the learning phase. One possible approach is data sanitisation, which involves filtering and erasing potentially poisonous samples from the learning dataset by identifying distinct characteristics or detecting anomalous patterns indicative of backdoor attacks~\cite{10.5555/3327757.3327896, aaai2019_activation_clustering, 263780, pmlr-v139-hayase21a, 9710118}. Another approach is robust learning, which neutralises the impact of backdoors by introducing randomness during the learning process. For example, adding random transformations to the learning data inflicts perturbations to trigger patterns~\cite{9414862} (e.g. cut-and-paste data augmentation~\cite{Yun_2019_ICCV}). Regularising gradients and adding random noises in the optimisation process may also enhance robustness~\cite{48698, Du2020Robust, DBLP:conf/ndss/NaseriHC22} (e.g. differential privacy~\cite{10.1145/2976749.2978318}). Ensemble learning trains a diverse collection of base models with randomised subsets of samples and aggregates the predictions of ensemble models for making inference, assuming that a majority of the base models are unlikely influenced by a minor amount of poisonous data~\cite{DBLP:conf/iclr/0001F21, Jia_Cao_Gong_2021, Jia_Liu_Cao_Gong_2022} (e.g. bootstrap aggregating~\cite{Breiman:1996aa}).

\subsubsection*{Reactive Paradigm}
Reactive defence counteracts the presence of backdoors by filtering or purifying either the samples or the models. Malicious samples can be eliminated by monitoring inputs for suspicious or anomalous patterns that could indicate a backdoor trigger or observing the predictions for unusual behaviour that may signal backdoor activation~\cite{10.1145/3359789.3359790, DBLP:journals/corr/abs-1912-01206, 10.1145/3400302.3415671}. These samples can also be purified by perturbing or reconstructing the poisonous regions~\cite{10.1145/3433210.3453108, 10.1145/3427228.3427264, 9754562}. Randomised smoothing can also be viewed as a form of purification, as it adds random noise to the samples to overwhelm injected triggers and makes predictions based on a majority vote over multiple noisy versions of each sample~\cite{wang2020certifying, 10.5555/3524938.3525700, 10179451}. Models with Trojans can be detected by constructing a meta-classifier and rejected for deployment if they are determined to be infected~\cite{9157782, 9519467, 10.1007/978-3-030-58583-9_20, 10.1007/978-3-030-58592-1_14, NEURIPS2021_8fd7f981, xiang2022posttraining}. These models can also be renovated with catastrophic forgetting~\cite{doi:10.1073/pnas.1611835114, 8119189, zeng2022adversarial}, knowledge distillation~\cite{44873, 10.1145/3411508.3421375, li2021neural} and neurone pruning~\cite{23864, Liu:2018aa, wu2021adversarial}.
 
\subsection{Problem Statement}
\subsubsection*{Context and Scope}
Based on the background information, we consider a common backdoor attack scenario in which the triggers are applied in cyberspace (space), cause an all-to-one mapping (causality), are generic for each sample (genericity), are arbitrary handcrafted patterns (optimality), represent semantically independent parts of samples (semanticity), and are visible to human perceptual systems (visibility). Furthermore, it is assumed that the attacker has bypassed automated detection and implanted backdoors into a neural network model during the learning phase (reactive paradigm). In particular, the threat model considered in this study corresponds to a classical backdoor attack scenario (known as BadNet~\cite{8685687}), as illustrated in Figure~\ref{fig:attack}. Let $\mathcal{D}$ and $\mathcal{D}'$ denote the sets of benign samples $(x, y)$ and poisoned samples $(x', y')$, respectively, where $x$ and $x'$ are input samples paired with their corresponding true labels $y$ and malicious target labels $y'$. The poisoned samples typically constitute a small proportion of the joint learning set (e.g., 5\%). The model is trained on the combined set to minimise a composite loss function, formulated as:
\begin{equation}
	\min \sum_{(x, y) \in \mathcal{D}} \mathcal{L}(f(x), y) + \sum_{(x', y') \in \mathcal{D}'} \mathcal{L}(f(x'), y') ,
\end{equation}
where $\mathcal{L}$ is the task-specific loss function such as cross-entropy.

On the defence side, we consider a scenario where the original learning dataset is no longer accessible. Access to the dataset may be restricted for the following reasons: to prevent potential misuse or breaches that could compromise sensitive information and individuals' privacy (privacy regulations); to protect the intellectual property and competitive advantages of companies or organisations (proprietary restrictions); due to difficulties and time-consuming retrieval methods associated with archiving and storing (archival policies); because of unsupported formats or incompatible systems (technical barriers); and because the dataset may be outdated, no longer maintained, or otherwise difficult to access (digital obsolescence). Hence, the possibility of uncovering hidden triggers by inspecting the dataset is restricted. 

\subsubsection*{Objective and Constraints}
This study focuses on neural networks used in image classification tasks. We assume the scenario where exact trigger content may be elusive but constraints on its size are available. In other words, we do not have precise information about what the trigger looks like, but the range within which its dimension falls. In practice, trigger dimensions are typically large enough to have a notable effect but small enough to evade detection. The research objective is to remove backdoors from a potentially infected model while maintaining its functionality. Although the original dataset is unavailable, we assume that a small amount of data sampled from the same or similar distribution is acquirable for analysing and unlearning the backdoors. Formally, we are given the following components:
\begin{itemize}
    \item A pre-trained image classification model $f: \mathcal{X} \to \mathcal{Y}$, where $\mathcal{X} \subseteq \mathbb{R}^n$ represents the input space (i.e., the space of images represented as $n$-dimensional vectors), and $\mathcal{Y}$ is the set of possible classes. Note that the original training dataset used to train $f$ is no longer available.
    
    \item A set of candidate masks $\mathcal{M}$ characterises potential backdoor triggers, where each mask $\boldsymbol{m} \in \mathcal{M}$ is constrained by a set of conditions, such as maximum allowed size or dimensions within an image.
        
    \item A clean dataset $\mathcal{D} = \{ (\boldsymbol{x}, y) \mid \boldsymbol{x} \in \mathcal{X}, y \in \mathcal{Y} \}$, serving both as a normative set for hypothesis analysis and as an unlearning set for machine unlearning, contains instances that are free from backdoor contamination, but comprises a much smaller number of instances than the original training set of $f$.
\end{itemize}
Let $\boldsymbol{x}'$ represent a malicious input generated by applying a backdoor trigger to a clean image $\boldsymbol{x} \in \mathcal{X}$ with a target class $y'$. We seek to find a modified classifier $f^*: \mathcal{X} \to \mathcal{Y}$  that minimises backdoor vulnerability (i.e. the probability that a malicious input is misclassified by $f^*$ as the target class)
\begin{equation} 
P(f^*(\boldsymbol{x}') = y') , 
\end{equation}
while ensuring knowledge fidelity (i.e. the probability that $f^*$ assigns the same classification to a benign input as $f$)
\begin{equation} 
P(f^*(\boldsymbol{x}) = f(\boldsymbol{x})). 
\end{equation}
To assess whether a model is infected and to establish the necessary hyper-parameters for this assessment, some aspect of the model's behaviour must be known \textit{a priori}. This is because understanding how the model should behave under normal conditions helps in identifying deviations that may indicate infection or compromise. This prior knowledge can be anticipated based on a surrogate model or derived from empirical evidence.

\subsection{Reverse Engineering}
A key aspect of this study is dedicated to reverse-engineering the trigger within the context of the specified attack and defence scenarios. The related research on this topic is briefly described as follows~\cite{8835365, 9338311, 9879000, wang2023unicorn}. Let $y$ denote a given target label to be analysed, $\boldsymbol{x}$ denote a data sample drawn from a set $\mathcal{X}$ and $f(\cdot)$ denote an infected classification model. To reverse-engineer the most likely backdoor trigger which would cause samples to be classified as the target label $y$, a common method involves solving the following optimisation problem consisting of a loss function $\mathcal{L}$ and a regularisation function $R$ weighted by a hyper-parameter $\lambda$:
\begin{equation}
\underset{ \{ \boldsymbol{z}, \boldsymbol{m} \}}{\operatorname{arg\,min}}\, \sum_{\boldsymbol{x} \in \mathcal{X}} \mathcal{L}(y, f((1-\boldsymbol{m})\boldsymbol{x} + \boldsymbol{m}\boldsymbol{z})) + \lambda \mathcal{R}(\boldsymbol{z}, \boldsymbol{m}),
\end{equation}
where the term inside $f(\cdot)$ denotes a manipulated sample, created by overwriting a potential trigger $\boldsymbol{z}$ onto a benign sample $\boldsymbol{x}$ using a mask $\boldsymbol{m}$. This optimisation process finds a pair of $\boldsymbol{z}$ and $\boldsymbol{m}$ that misleads classification (evaluated by $\mathcal{L}$) and satisfies certain prior assumptions, empirical knowledge or practical heuristics (regularised by $\mathcal{R}$). The possible regularisation terms include, but are not limited to, the $L_{p}$ norm, which restricts the size and magnitude of the solutions, as well as the total-variation norm, which encourages smooth solutions. Then, an outlier detection is applied to identify the malicious trigger from all the potential ones generated from the optimisation process.

\section{Conceptual Framework}
In this section, we briefly explain the core rationales built into our conceptual framework and illustrate how interdisciplinary concepts are related, providing an overview of the theoretical foundation.

\subsection{Hypnopaedia}
In a metaphorical sense, a backdoor attack can be considered a form of mind-hacking that indoctrinates or implants an idea into a machine's subconscious mind. A psychological reminiscence for backdoors is \emph{hypnopaedia}, which refers to learning under the state of hypnosis or unconsciousness, conditioning an individual's beliefs and behaviours without their conscious awareness~\cite{huxley1932}. A backdoor trigger is analogous to a \emph{hypnotic suggestion} used to subject an individual undergoing hypnosis to the command of a hypnotist.

\subsection{Cybernetics}
This study applies cybernetic principles to manage the risk of backdoors arising from dynamic data sources. Cybernetics is the study of automation with an emphasis on circular causality and regulatory feedback for controlling systems automatically~\cite{ashby1963introduction}. Feedback loops are fundamental to cybernetics because they enable systems to self-regulate and react to changes in their environment (\emph{external dynamics}) or within themselves (\emph{internal dynamics}). Let us take a thermostat as an example to demonstrate how cybernetic principles are applied in a simple feedback control system~\cite{10.7551/mitpress/11810.001.0001}. A thermostat operates continuously, monitoring the temperature of a room (the controlled variable) and reacting to changes in the environment. Its goal is to maintain a stable temperature around a set-point. It contains a \emph{sensor} that detects the current temperature and a \emph{controller} that governs whether to turn on or off the heating or cooling system based on a desired set-point. It then sends control signals to an \emph{actuator} to adjust the temperature accordingly.

\subsection{Metacognition}
Analogously, a backdoor-aware learning machine can be modelled as a cybernetic system. A learning machine (or its state and parameters) is analogous to the temperature of a room, which is the variable being controlled. A learner updates the machine constantly to adapt to continuous streams of new information and reports the changed state, acting like a sensor which observes and measures external dynamics from an ever-changing real world. A controller evaluates whether there are any backdoors present in the current state and controls whether actions need to be taken to address detected backdoors. It empowers a machine with \emph{metacognition}, referring to the awareness of one's own cognition and knowledge, thereby allowing a machine to analyse and monitor its own thinking patterns~\cite{Cope:2022aa}. An unlearner functions like an actuator that either reacts to the detected backdoors or maintains the current state based on the control decision. If a reaction is needed, it updates the machine with an unlearning set of samples (alongside other auxiliary knowledge) to remove potential backdoors.

\subsection{Motivated Forgetting}
Machine unlearning parallels a psychological phenomenon of \emph{motivated forgetting}, where people suppress or repress unwanted memories consciously or unconsciously~\cite{https://doi.org/10.1111/j.1467-6494.1968.tb01470.x}. This can occur due to the desire to suppress unpleasant memories or reduce cognitive dissonance. Similarly, machine unlearning describes the process where a machine forgets or adjusts its learned patterns and associations~\cite{9519428}. This is often necessary when the model has learned something undesirable, inaccurate or outdated. If the backdoor trigger is estimated through reverse engineering, the machine can be fine-tuned to disassociate its behaviours from the estimated trigger, thereby reducing the influence of the backdoor.

\subsection{Memory Retrieval}
Nonetheless, trigger estimation can be challenging since there is a vast amount of potential trigger patterns and target behaviours. Since the backdoor is typically introduced during the learning phase, a logical solution is to inspect the learning dataset. However, the original dataset is often inaccessible due to constraints such as privacy regulations, proprietary restrictions, archival policies, technical barriers, and digital obsolescence. As an alternative, one approach is to extract information directly from a machine’s memory. That is, model inversion is a \emph{memory retrieval} technique that reverse-engineers a model to infer information about its learning dataset~\cite{10.1145/2810103.2813677}. This can be achieved by submitting queries to a model iteratively and adjusting the query based on its response, finding the optimal query that maximising the activation through trial and error. In investigative psychology, there is a similar technique used by law enforcement during criminal investigations to retrieve information about a crime scene from eyewitnesses, referred to as \emph{cognitive interview}~\cite{Geiselman:1986aa}. It involves multiple questioning techniques and mnemonic strategies to facilitate the mental process of recall or recollection, eliciting memories associated with a specific event from the past.

\subsection{Butterfly Effect}
In the context of memory retrieval, an individual's recall might stabilise around certain dominant narratives or repeated rehearsals. This phenomenon may also occur in model inversion, where the outcomes consistently return to a set of convergent but potentially suboptimal patterns. This limited set of patterns can be thought of as an \emph{attractor} in a subject's memory. In chaos theory, an attractor is a cluster of states towards which a system tends to evolve, regardless of small variations in initial conditions. In essence, to escape or diverge from an attractor means disrupting the stability of the system, making it more sensitive to initial conditions. This concept is reminiscent of the \emph{butterfly effect}, which illustrates how small changes in initial conditions can significantly influence a dynamic system's orbital trajectory, leading to vastly diverse outcomes. In cognitive interview, recall can be constrained by the wording of the questions, which acts as an attractor in chaos theory influences the trajectory of a system, guiding it towards certain states or behaviours. Varied prompts and diverse questions may then be used to diverge from this attractor, encouraging broader and more accurate recall. In model inversion, divergence from attractors can be encouraged by introducing stochastic processes.

\subsection{Psychometrics}
The outcomes of model inversion can be viewed as representational content that reflects the internalised knowledge of a model, reconstituted in a form that resembles the learning set of samples (or projected back to the sample space). These outcomes resemble \emph{mental imagery} in the human mind, serving as a conceptual representation of things and experiences~\cite{GALTON:1880aa, sep-mental-representation, Edelman:1995aa}. The objective is to identify a potentially malicious trigger pattern within the realm of the mind's visual representations, analogous to psychometric assessment of an individual's potential for criminality. Such a pattern could induce \emph{sensory deprivation} and stimulate \emph{hallucinations}, distorting the perception of a machine~\cite{doi:10.1126/science.133.3467.1808}. In other words, such hallucinatory patterns can manifest through backdoor activation that consistently diverts the machine's behaviours from expected outcomes. Therefore, the likelihood of a hypothetical pattern being the actual trigger and the probability of infection can be quantified through activation statistics.

\begin{figure*}[!t]
\centering
\includegraphics[width=0.98\linewidth]{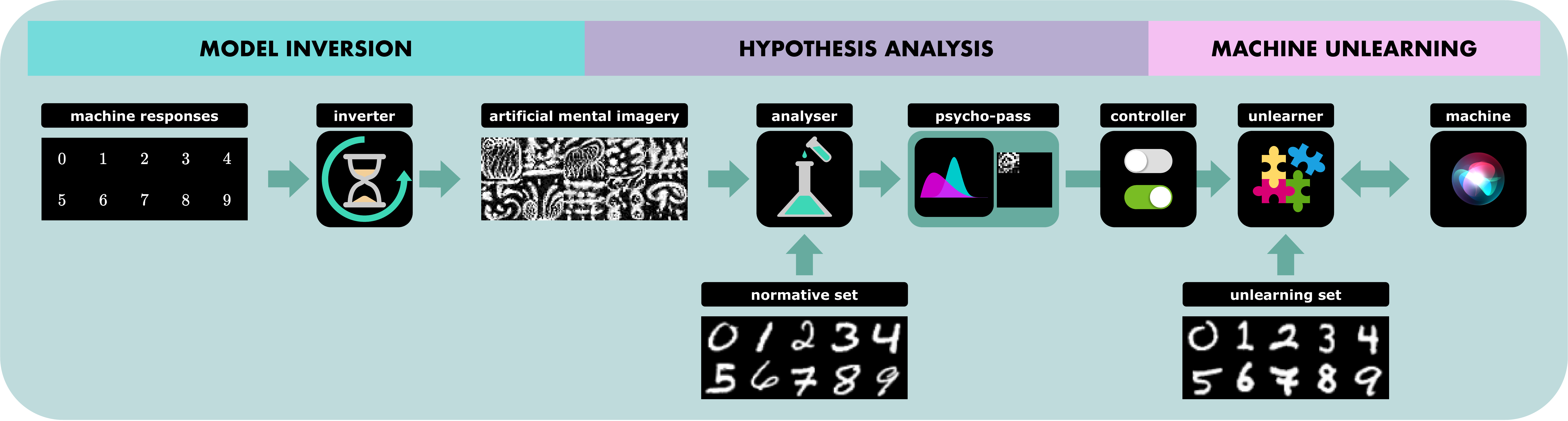}
\caption{Systematic defence against neural backdoors via model inversion, hypothesis analysis and machine unlearning.}
\label{fig:defence}
\end{figure*}

\section{Methodology}
Our proposed method consists of three parts: model inversion, hypothesis analysis and machine unlearning. Model inversion retrieves artificial mental images that represent prototypes for all possible classes of samples. Hypothesis analysis quantifies the likelihood of each hypothetical pattern drawn from the artificial mental images being the actual trigger with the aid of a \emph{normative set} of data, and estimates the probability that a machine is infected. Machine unlearning applies the most likely trigger pattern on an \emph{unlearning set} of data to disassociate it from the conditioned response. Both normative and unlearning sets of data contain a small number of correctly labelled samples and may overlap. An overview of the proposed method is outlined in Figure~\ref{fig:defence}.

\subsection{Model Inversion}
Model inversion aims to invert a machine-learning model to infer the information about its learning data. The objective is to find an $n$-dimensional synthetic input $\boldsymbol{z}_y \in \mathbb{R}^{n}$ that minimises the discrepancy between the output of the model $f(\boldsymbol{z}_y)$ and a target output $y$. This can be metaphorically seen as retrieving an artificial mental image about a particular object in a machine's memory. Mathematically, this can be formulated as finding the optimal synthetic input for an unconstrained minimisation problem:
\begin{equation}
\underset{\boldsymbol{z}_y}{\operatorname{arg\,min}}\, \mathcal{L}(y, f(\boldsymbol{z}_y)),
\end{equation}
where $\mathcal{L}$ denotes a loss function measuring the discrepancy between the ground-truth response and the prediction. For multinomial classification, the loss is usually calculated using cross-entropy or negative log-likelihood. The model inversion process is outlined in Algorithm~\ref{alg:inverse}.

\begin{figure*}[!t]
\centering
\includegraphics[width=0.98\linewidth]{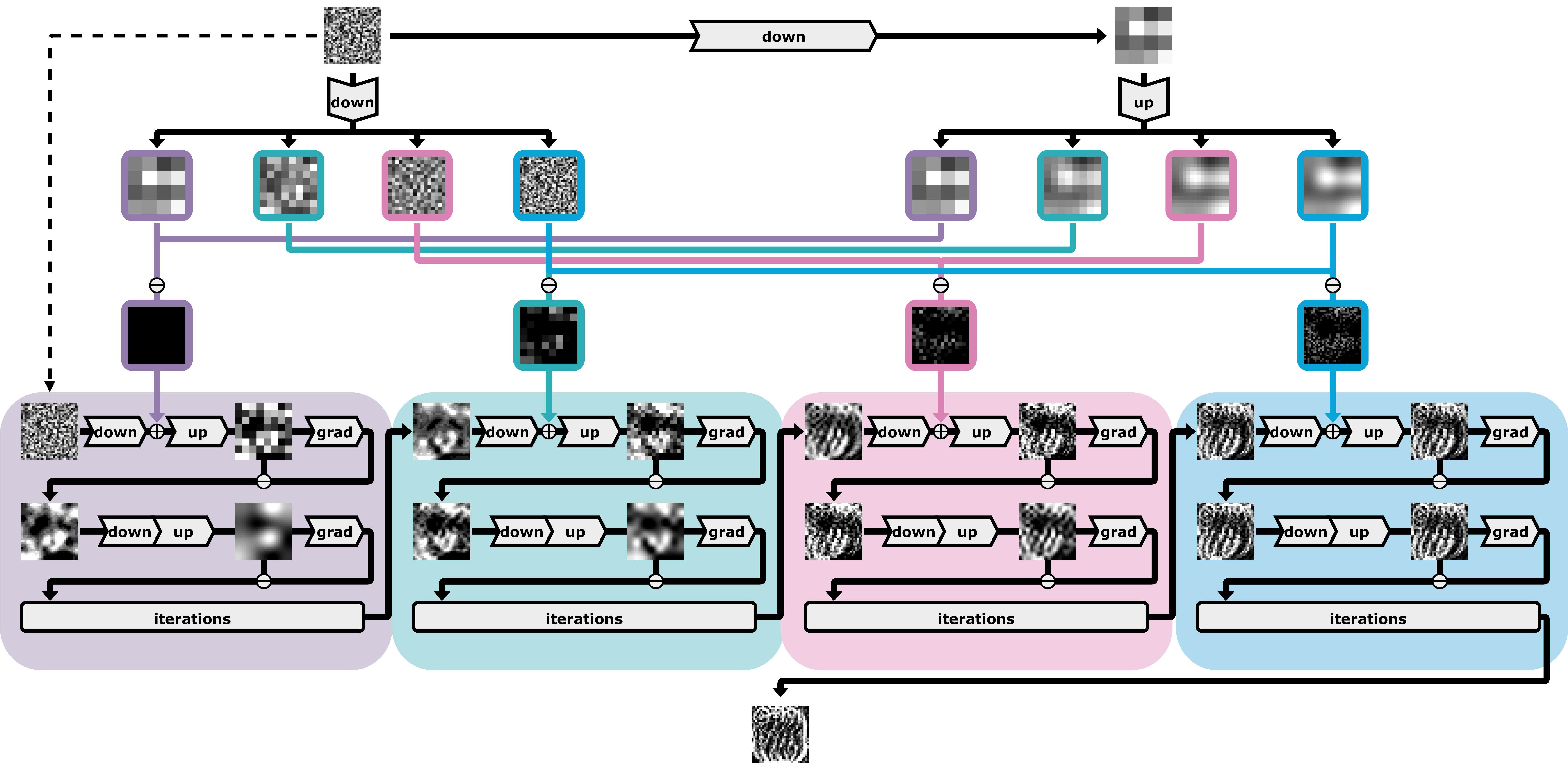}
\caption{Illustration of multi-scale model inversion for projecting an artificial mental image with a random initial noise.}
\label{fig:inversion}
\end{figure*}

\subsubsection*{Gradient Descent}
Gradient descent, a first-order optimisation algorithm, offers a principled approach to model inversion by iteratively adjusting and updating the input data in the direction that minimises a given loss function. Let $\boldsymbol{z}_y^{0}$ be randomly selected values (as an initial guess) of an input sample. For each iteration, the sample is updated by
\begin{equation}
\boldsymbol{z}_y^{(t)} = \boldsymbol{z}_y^{(t-1)} - \delta \cdot \operatorname{sgn}( \nabla_{\boldsymbol{z}_y} \mathcal{L}(y, f(\boldsymbol{z}_y^{(t-1)})) ),
\end{equation}
where $\delta$ is the step size and the subsequent term is the sign of the gradient of the loss function with respect to the input $\boldsymbol{z}_y$ evaluated at $\boldsymbol{z}_y^{(t-1)}$. The iterative update is repeated until a convergence criterion is met. This criterion can be a maximum number of iterations, reaching a threshold value of the loss function, or observing negligible changes in $\boldsymbol{z}_y$ between iterations. Once the convergence criterion is satisfied, the final value $\boldsymbol{z}_y$ represents an artificial mental image for which the prediction of the model $f(\boldsymbol{z}_y)$ approximates the target response $y$. An artificial mental image can be viewed as the centroid of samples belonging to a particular class. This lies in the fact that the model may learn to recognise a class by essentially memorising the average or typical features within that class. In practice, we may synthesise multiple images for each class with different random initial states, rather than a single image, to increase the likelihood of successfully unveiling backdoor triggers.

%%%
\subsubsection*{Multi-Scale Optimisation}
Model inversion behaves as a deterministic function given the initial conditions. In practice, while inputs are randomly initiated, the outputs often converge to similar patterns. This phenomenon can be viewed, by analogy, as related to the concept of attractors in chaos theory. An attractor is a set of states that a dynamic system naturally moves toward over time, despite minor variations in its initial state. Divergence from such orbital trajectory can be encouraged by introducing probabilistic or stochastic processes. To implement this concept heuristically in the context of model inversion, a multi-scale optimisation technique is developed for progressively refining artificial mental images at various resolutions with the stochastic number of iterations for each scale, as depicted in Figure~\ref{fig:inversion}. Note that when the number of iterations for each intermediate scale is randomised as zero, multi-scale optimisation degenerates into single-scale optimisation. Initially, the optimisation process is operated at a small spatial resolution, identifying macro changes that guide the model to interpret the image as a specific target output. Following the completion of optimisation at the current scale, the image is upsampled to the next resolution with the addition of resampling residuals, compensating for the information loss due to resampling. Let $\boldsymbol{z}_y^{\text{max}}$ denote the initial random guess at the maximum resolution and $\boldsymbol{z}_y^{\text{min}}$ its counterpart at the minimum resolution. The resampling residuals $\boldsymbol{\rho}$ represent the information loss between the downsampled version of $\boldsymbol{z}_y^{\text{max}}$ and the upsampled version of $\boldsymbol{z}_y^{\text{min}}$ at a certain resolution, as computed by
\begin{equation}
\boldsymbol{\rho} = \operatorname{resample}_{\downarrow} (\boldsymbol{z}_y^{\text{max}}) - \operatorname{resample}_{\uparrow} (\boldsymbol{z}_y^{\text{min}}).
\end{equation}
These residuals are added to the intermediate results at the beginning of each resolution-wise optimisation process to offset the information loss caused by resampling. The progressive optimisation process then continues to capture finer details until reaching the final resolution.

\subsubsection*{Adversarial Learning}
An effective unlearning of backdoors relies largely on the quality of artificial mental images generated by model inversion. However, the complexity and variability of data make model inversion more challenging, compared to the simpler and more consistent data. This leads to inferior inversion results for complex datasets due to the difficulties in accurately capturing and reconstructing the intricate textures and diverse features present in such data. As a consequence, while inversion on simple dataset may yield clear and recognisable synthetic content, inversion on complex dataset often produces blurry and less interpretable results. Adversarial learning enhances the robustness and clarity of latent representations learned by machine learning models, which may translate to better performance in model inversion, yielding clearer and more interpretable synthetic content. This occurs because robust representations focus on essential and discriminative aspects of the data, reducing the impact of noise and irrelevant details, thereby leading to more accurate and visually distinct reconstructions. It involves incorporating adversarial examples into the learning process~\cite{DBLP:journals/corr/SzegedyZSBEGF13, 43405, kurakin2017adversarial, 10.5555/3454287.3454814}. A common method for generating adversarial examples is projected gradient decent (or ascent), which iteratively applies small perturbations and projects perturbed examples back into a valid sample space~\cite{Madry:2018aa}. It moves a sample towards the direction that maximally increases the loss and thereby increases the likelihood of causing the model to misclassify the perturbed sample. In practice, to train a model on a mixture of perturbations with varying levels of intensity, we randomly sample the maximum number of iteration steps for each instance. An adversarial example to be generated at an iteration step $t$ is given by
\begin{equation}
	\boldsymbol{x}_{\text{adv}}^{(t)} = \operatorname{proj}_{\epsilon} ( \boldsymbol{x}_{\text{adv}}^{(t-1)} + \alpha \cdot \nabla_{\boldsymbol{x}_{\text{adv}}} \mathcal{L}(y, f(\boldsymbol{x}_{\text{adv}}^{(t-1)})) ),
\end{equation}
where $\alpha$ denotes a step size and $\operatorname{proj}_{\epsilon}$ denotes a projection function that regularises the maximum perturbation magnitude with a threshold parameter $\epsilon$. For instance, to project a sample the onto an $L_{\infty}$ ball of radius $\epsilon$ centred at the initial state, we truncate the perturbations that excess $\epsilon$. This ensures the distortion bounded within the given constraint.

% ------- Algorithm: Model Inversion -------
\begin{algorithm}[t]
\caption{Model Inversion}\label{alg:inverse}
\begin{algorithmic}
\State \textbf{Input}: class label $y$
\State \textbf{Output}: artificial mental image $\boldsymbol{z}_y$
\State
\LineComment{\textit{initialisation}}
\State set scale factors
\State set step size $\delta$
\State initialise randomly $\boldsymbol{z}_y$
\State compute $\boldsymbol{z}_y^{\text{max}}$ and $\boldsymbol{z}_y^{\text{min}}$ as max-scale and min-scale $\boldsymbol{z}_y$
\State
\LineComment{\textit{multi-scale gradient-descent optimisation}}
\For {each scale factor}
\State resample $\boldsymbol{z}_y$ by current scale factor
\State compute $\boldsymbol{\rho} \gets \operatorname{resample}_{\downarrow}(\boldsymbol{z}_y^{\text{max}}) - \operatorname{resample}_{\uparrow}(\boldsymbol{z}_y^{\text{min}})$
\State update $\boldsymbol{z}_y \gets \boldsymbol{z}_y + \boldsymbol{\rho}$
\While {convergence criterion is not satisfied}
\State upsample $\boldsymbol{z}_y$ for gradient computation
\State compute gradient $\nabla_{\boldsymbol{z}_y} \mathcal{L}(y, f(\boldsymbol{z}_y))$
\State update $\boldsymbol{z}_y \leftarrow \boldsymbol{z}_y - \delta \cdot \operatorname{sgn}( \nabla_{\boldsymbol{z}_y} \mathcal{L}(y, f(\boldsymbol{z}_y)) )$
\State downsample $\boldsymbol{z}_y$ by current scale factor
\EndWhile
\EndFor
\end{algorithmic}
\label{alg1}
\end{algorithm}

\subsection{Hypothesis Analysis}

Suppose a set of artificial mental images $\mathcal{Z}$ is generated via model inversion. For each class $y$, multiple artificial mental images $\boldsymbol{z}_y$ may be generated through repeated inversion trials. Typically, each image reflects the features of samples belonging to a certain class $y \in \mathcal{Y}$. For the target backdoor class, the features of the trigger may also manifest themselves in the corresponding image. The likelihood that a hypothetical pattern reflects the feature of the actual trigger can be quantified by assessing its impact on a set of normative data. With some prior knowledge acquired from historical data, we can further infer the probability that the current machine is in an infected state based on the likelihood of the most plausible hypothesis. The hypothesis analysis process is outlined in Algorithm~\ref{alg:h_analysis}.

\begin{figure}[!t]
\centering
\includegraphics[width=0.99\linewidth]{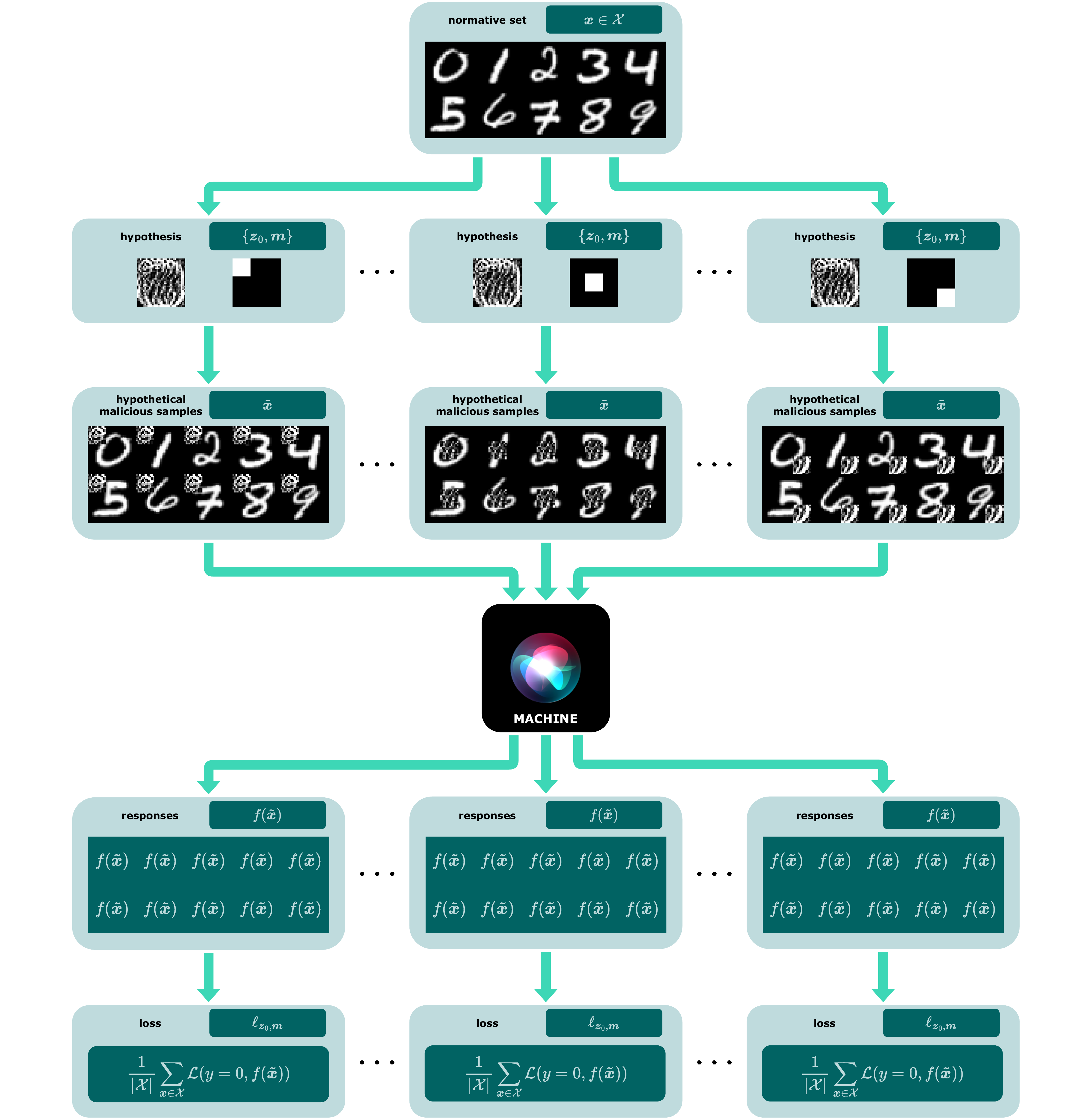}
\caption{Illustration of hypothesis analysis process for an artificial mental image and a series of attention masks given a normative set of samples.}
\label{fig:hypothesis}
\end{figure}

\subsubsection*{Maximum Likelihood Estimation}
Let $\boldsymbol{x} \in \mathcal{X}$ be a clean or benign sample from a normative set and $\boldsymbol{m} \in \mathcal{M}$ be a hypothetical mask from a candidate mask set. In practice, we may generate a set of masks with a sliding window of a customised size. A hypothetically malicious or toxic sample can be created by
\begin{equation}
	\tilde{\boldsymbol{x}} = (1-\boldsymbol{m})\boldsymbol{x} + \boldsymbol{m}\boldsymbol{z}_y,
\end{equation}
where a pair $\{\boldsymbol{z}_y, \boldsymbol{m}\}$ represents a hypothetical trigger with the assumption that the class $y$ (associated with $\boldsymbol{z}_y$) is the backdoor class. Theoretically, a hypothetical trigger that resembles the actual trigger (either visually or abstractly) would cause samples classified as a certain target class, implying backdoor activation. Let $\mathcal{H}$ denote the set of hypotheses, defined as the Cartesian product of $\mathcal{Z}$ and $\mathcal{M}$. For a hypothesis $h$ consisting of a pair $\{\boldsymbol{z}_y, \boldsymbol{m}\}$, the likelihood of this hypothesis representing the actual trigger can be evaluated by considering how well the hypothetical trigger leads to the hypothetical target class when passed through the model. Specifically, the likelihood is inversely proportional to the average loss computed by comparing the predictions on hypothetically malicious samples with a hypothetical target class:
\begin{equation}
	\ell_{h} = \frac{1}{| \mathcal{X} |} \sum_{\boldsymbol{x} \in \mathcal{X}} \mathcal{L}(y, f(\tilde{\boldsymbol{x}})),
\end{equation}
In essence, a lower average loss indicates that the hypothesis biases the model's behaviours more significantly, leading to a higher likelihood of representing the actual trigger. The most likely hypothesis is the one that yields the minimum average loss, as given by
\begin{equation}
	h^{*}\text{: } \{ \boldsymbol{z}_y^{*}, \boldsymbol{m}^{*} \} = \underset{ h \in \mathcal{H} }{\operatorname{arg\,min}}\, \frac{1}{| \mathcal{X} |} \sum_{\boldsymbol{x} \in \mathcal{X}} \mathcal{L}(y, f(\tilde{\boldsymbol{x}})) .
\end{equation}

\begin{figure}[!t]
\centering
\includegraphics[width=0.95\linewidth]{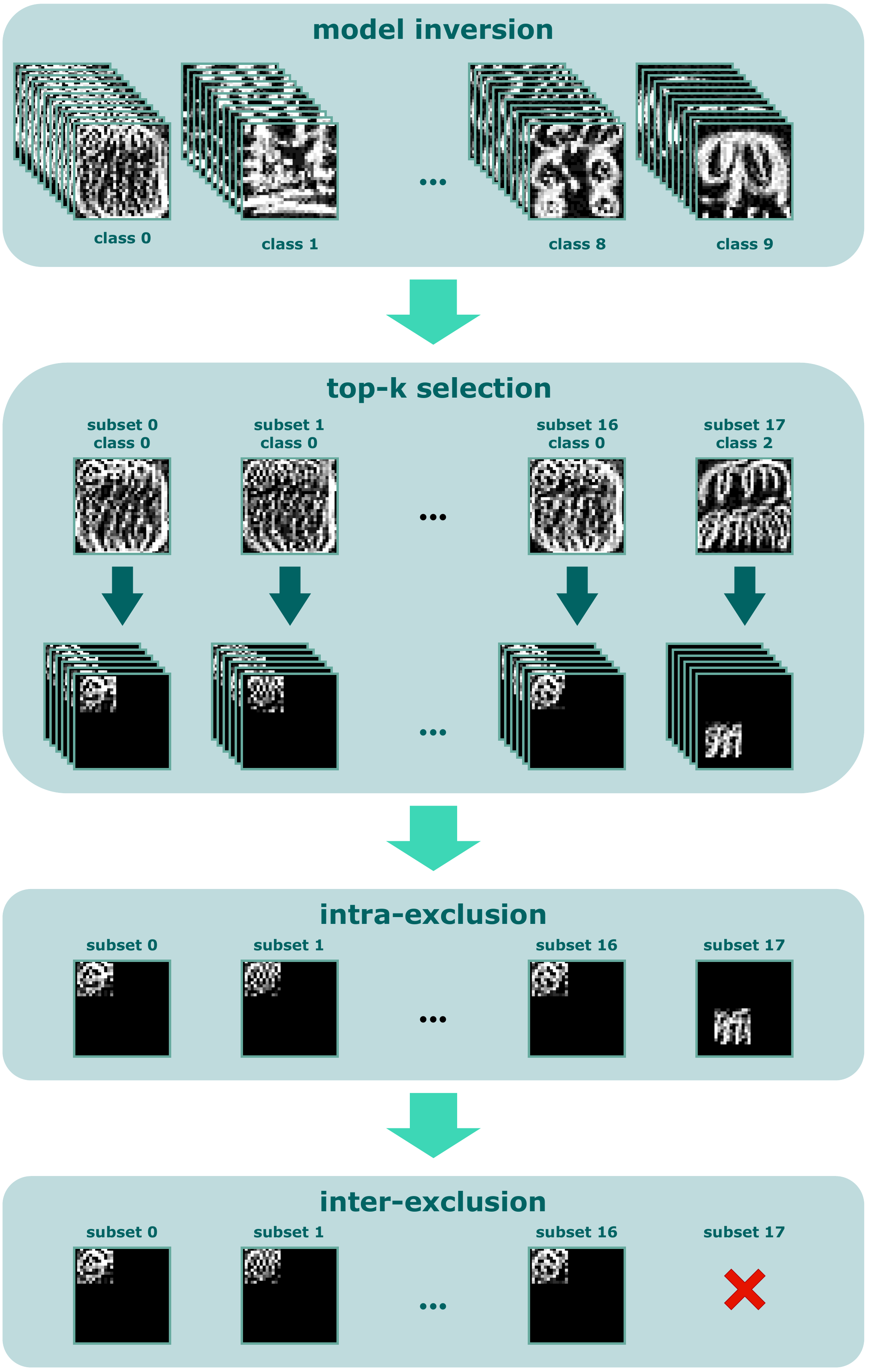}
\caption{Illustration of outlier exclusion process for eliminating intrinsic natural triggers and retaining extrinsic artificial triggers.}
\label{fig:outlier}
\end{figure}

% ------- Algorithm: Hypothesis Analysis -------
\begin{algorithm}[t]
\caption{Hypothesis Analysis}\label{alg:h_analysis}
\begin{algorithmic}
\State \textbf{Input}: normative set $\mathcal{D}\text{: }\{\mathcal{X}, \mathcal{Y}\}$ and mental image set $\mathcal{Z}$
\State \textbf{Output}: hypothesis set $\mathcal{H}^*$ and posterior $P(s_1 | e)$
\State
\LineComment{\textit{initialisation}}
\State set candidate mask set $\mathcal{M}$
\State define $\mathcal{H}\text{: } \mathcal{Z} \times \mathcal{M}$
\State initiate a sequence of $\ell_{h}$ where $h = {\{\boldsymbol{z}, \boldsymbol{m}\} \in \mathcal{H}}$

\State
\LineComment{\textit{maximum likelihood estimation}}
\For {$\boldsymbol{z}_y \in \mathcal{Z}$}
\State get the corresponding class label $y \in \mathcal{Y}$
\For {$\boldsymbol{m} \in \mathcal{M}$}
\State compute total loss $\ell_{h} \gets \sum_{\boldsymbol{x}\in \mathcal{X}} \mathcal{L}(y, f(\tilde{\boldsymbol{x}})$
\State where $\tilde{\boldsymbol{x}} \gets (\boldsymbol{1} - \boldsymbol{m})\boldsymbol{x} + \boldsymbol{m}\boldsymbol{z}_y$
\EndFor
\EndFor

\State
\LineComment{\textit{outlier exclusion}}
\State select the top $k$ hypotheses with the minimum losses
\State do intra-exclusion to get cluster centroids
\State do inter-exclusion to get homogeneous cluster centroids
\State update hypothesis set $\mathcal{H}^*$

\State
\LineComment{\textit{Bayesian inference}}
\State compute the evidence $e$
\State compute priors $P(s_0)$ and $P(s_1)$
\State compute likelihoods $P(e \vert s_0)$ and $P(e \vert s_1)$
\State compute marginal likelihood $P(e) = \sum_{s_i} P(e \vert s_i) P(s_i)$
\State infer posterior $P(s_1 \vert e)$
\end{algorithmic}
\label{alg1}
\end{algorithm}

\subsubsection*{Outlier Exclusion}
In practice, the true backdoor hypothesis may yield a small, though not necessarily the minimum, average inversion loss due to the presence of deceptive outlier patterns. These outliers may arise when a small pattern, despite its limited size, carries sufficient hallucinatory features to mislead multiple inputs toward the same class. Such patterns can act as \emph{intrinsic natural triggers}, as opposed to \emph{extrinsic artificial triggers} injected during poisoning. To mitigate this effect, we implement an outlier exclusion procedure based on the empirical observation that genuine trigger patterns tend to emerge repeatedly across multiple inversion trials with both spatial and perceptual consistency, while spurious patterns do not exhibit such regularity. The outlier exclusion process consists of three stages: top-$k$ selection, intra-exclusion, and inter-exclusion, as illustrated in Figure~\ref{fig:outlier}.

\textit{Top-$k$ Selection}: 
Given a set of all hypotheses $\mathcal{H}$, where each hypothesis $h \in \mathcal{H}$ is associated with an average inversion loss $\ell_h$, rather than relying solely on the single hypothesis with the minimum loss, we retain a subset consisting of the $k$ hypotheses with the smallest loss values, such that
\begin{equation}
\mathcal{H}^{(k)} \subset \mathcal{H} .
\end{equation}

\textit{Intra-Exclusion}:
Let $\mathcal{H}_{\boldsymbol{z}_y}^{(k)}$ denote the subset of top-k hypotheses $\mathcal{H}^{(k)}$ that originates from the same artificial mental image $\boldsymbol{z}_y$:
\begin{equation}
\mathcal{H}_{\boldsymbol{z}_y}^{(k)} = \{ h \in \mathcal{H}^{(k)} \mid h = (\boldsymbol{z}_y, \boldsymbol{m}) \text{ for some } \boldsymbol{m} \} .
\end{equation}
For each subset $\mathcal{H}_{\boldsymbol{z}_y}^{(k)}$, we construct spatial clusters based on mutual proximity of the associated mask positions. In a cluster, every hypothesis $h = (\boldsymbol{z}_y, \boldsymbol{m})$ must have at least one neighbouring hypothesis $h' = (\boldsymbol{z}_y, \boldsymbol{m}')$ such that the spatial distance between their mask positions is within a radius $r$. Formally, two hypotheses $h$ and $h'$ are assigned to the same cluster if
\begin{equation}
| \operatorname{pos}(\boldsymbol{m}) - \operatorname{pos}(\boldsymbol{m}') | \leq r .
\end{equation}
Each cluster is represented by a single hypothesis, referred to as the cluster centroid, which is selected as the one with the minimum average loss:
\begin{equation}
c = \argmin_{h \in \text{cluster}} \ell_{h} ,
\end{equation}
thereby eliminating spatially redundant hypotheses that are considered geometric translations of the same pattern.

\textit{Inter-Exclusion}:
A cluster centroid $c$ is retained only if it has a sufficient number of homogeneous counterparts. Two cluster centroids $c$ and $c'$ are defined as homogeneous if they belong to the same class $y$ and their perceptual distance below a threshold $\delta$. Formally, the set of homogeneous centroids associated with $c$ is defined as:
\begin{equation}
\operatorname{homogene}(c) = \{c' \mid \operatorname{percep}(c, c') \leq \delta \}.
\end{equation}
The perceptual metric applied is the learned perceptual image patch similarity (LPIPS). A centroid $c$ is retained in the final hypothesis set $\mathcal{H}^*$ only if the number of its associated homogeneous cluster centroids are greater than a threshold $\theta$:
\begin{equation}
|\operatorname{homogene}(c)| > \theta .
\end{equation}
The rationale behind is that true trigger patterns tend to emerge consistently around similar spatial locations with perceptually similar appearances across multiple inversion trials, whereas spurious or outlier patterns tend to be more diverse and inconsistent.

The final set $\mathcal{H}^*$ contains the most reliable trigger hypotheses after excluding both spatially redundant and perceptually inconsistent patterns. The parameters involved in this outlier exclusion process include the number of selected patterns $k$, the radius for the neighbour patterns $r$, the threshold for perceptual similarity $\delta$ and the threshold for the number of homogeneous cluster centroids $\theta$.

\begin{figure}[!t]
\centering
\includegraphics[width=0.9\linewidth]{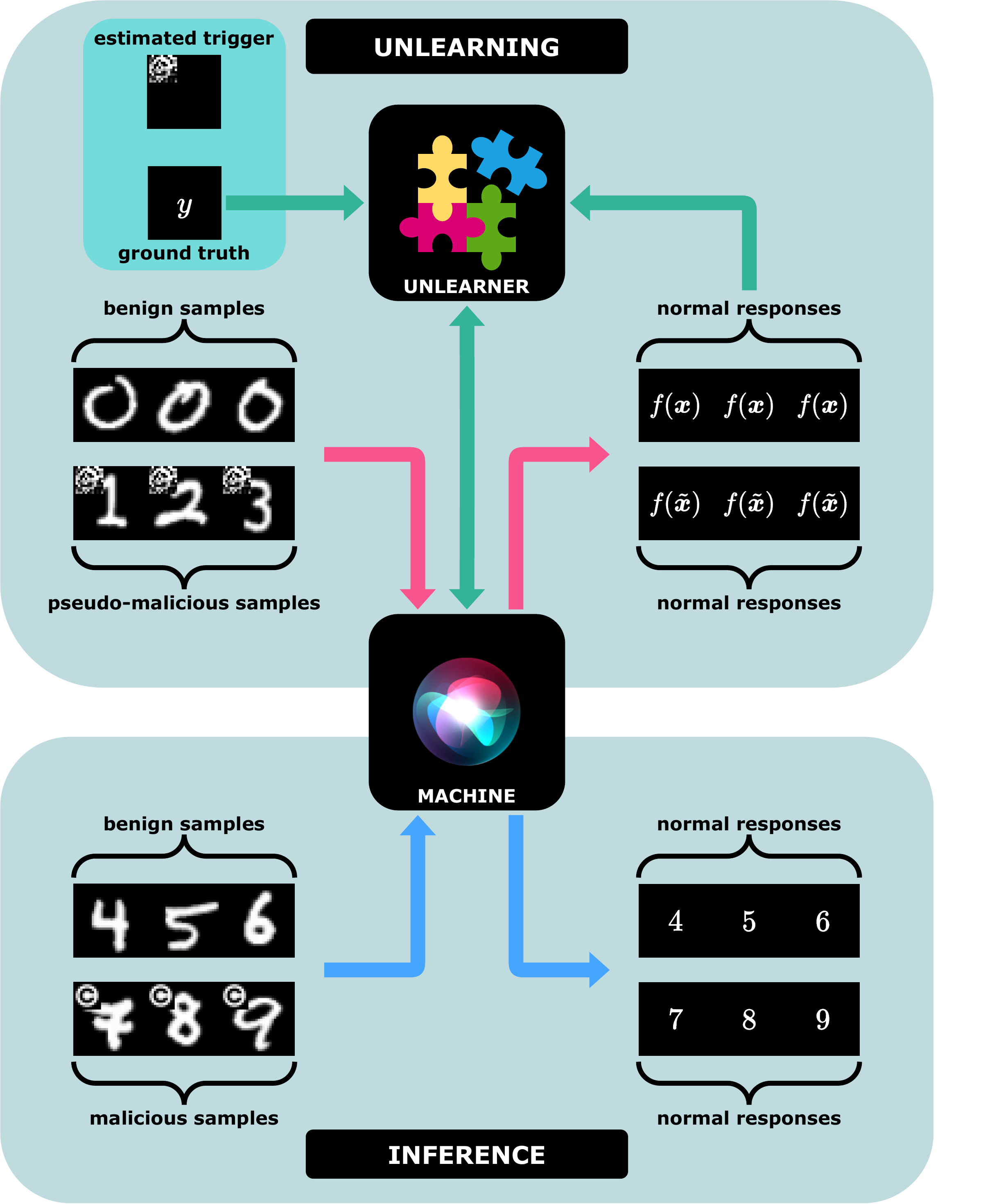}
\caption{Illustration of machine unlearning process.}
\label{fig:unlearning}
\end{figure}

\subsubsection*{Bayesian Inference}
To determine whether the machine should undergo the unlearning process, it is essential to infer the probability of a machine being in the uninfected state $s_0$ or the infected state $s_1$. If the machine is infected, each artificial mental image $\boldsymbol{z}_y$ from the selected hypotheses may exhibit a mixture of features from its corresponding class $y$ and the backdoor trigger. In contrast, if the machine is uninfected, these images are more likely to accurately reflect the intended class. This observation implies that, in uninfected models, the artificial mental images should be mutually consistent, whereas in infected models, the mental images of the target class may perceptually diverge from those generated by uninfected models. To quantify how such similarities or deviations indicate non-infection or infection, respectively, we rely on surrogate models to collect historical data and establish reference distributions. Suppose we have a small amount of independent and identically distributed (i.i.d.) auxiliary data available for training surrogate models. Using this data, we construct two surrogate models: one representing an uninfected machine, and the other representing an infected machine injected with an arbitrary trigger pattern. For a selected target class $\tilde{y}$, we generate artificial mental images from both surrogate models. These are denoted by $\mathcal{Z}_{\tilde{y}}^{(0)}$ for the uninfected surrogate and $\mathcal{Z}_{\tilde{y}}^{(1)}$ for the infected surrogate. We then compute perceptual distances in two ways. The first involves measuring distances within the set of artificial mental images generated by the uninfected surrogate model, forming the intra-model comparison set:
\begin{equation}
\mathcal{S}_0 = \left\{ \operatorname{percep}(z, z') \mid z \in \mathcal{Z}_{\tilde{y}}^{(0)}, z' \in \mathcal{Z}_{\tilde{y}}^{(0)} \right\} ,
\end{equation}
The second measures distances between the uninfected and infected surrogate models, forming the inter-model comparison set:
\begin{equation}
\mathcal{S}_1 = \left\{ \operatorname{percep}(z, z') \mid z \in \mathcal{Z}_{\tilde{y}}^{(0)}, z' \in \mathcal{Z}_{\tilde{y}}^{(1)} \right\} .
\end{equation}
These scores serve as historical data for probabilistic reasoning. For diagnosing a query machine, we derive the most probable hypotheses $\mathcal{H}^*$ and compute an average perceptual distance between between the corresponding artificial mental images $\mathcal{Z}_{y}^{(h)}$ and the reference set $z \in \mathcal{Z}_{y}^{(0)}$ from the uninfected surrogate:
\begin{equation}
e = \frac{1}{|\mathcal{Z}_{y}^{(h)}| \cdot |\mathcal{Z}_{y}^{(0)}|} \sum_{ z \in \mathcal{Z}_{y}^{(h)} }  \sum_{ z \in \mathcal{Z}_{y}^{(0)} }  \operatorname{percep}(z, z') .
\end{equation}
 This scalar value $e$ represents the observed evidence and is compared against the distributions of historical perceptual distances to infer the likelihood of infection. Bayesian inference is then used to compute the posterior probability from the beliefs and observed data. Specifically:
\begin{itemize}
	\item The prior probability $P(s_i)$ represents the initial belief about state $s_i$ (a discrete variable).
	\item The likelihood $P(e \vert s_i)$ represents the probability of observing evidence $e$ (a continuous variable) given that the machine is in state $s_i$.
	\item The marginal likelihood $P(e)$ represents the probability of observing evidence $e$ under all possible states, computed by integrating $P( e \vert s_i) P(s_i)$. 
\end{itemize}
 Applying Bayes' theorem, the posterior probability of the machine being infected given the observed evidence $e$ is given by
\begin{equation}
	\overbrace{P(s_1 \vert e)}^{\text{posterior}} = \frac{\overbrace{P(e \vert s_1)}^{\text{likelihood}} \overbrace{P(s_1)}^{\text{prior}}}{\underbrace{P( e \vert s_0) P(s_0) + P(e \vert s_1) P(s_1)}_{\text{marginal likelihood}}} .
\end{equation}
For simplicity, a \emph{non-informative prior} may be applied, assigning equal probability to each state, reflecting neutral prior knowledge and intending to have minimal influence on the posterior distribution. The likelihood functions $P(e \vert s_0)$ and $P(e \vert s_1)$ are estimated using kernel density estimation over the historical data $\mathcal{S}_0$ and $\mathcal{S}_1$, respectively. Kernel density estimation enables non-parametric estimation of probability densities without assuming any predefined distribution form. To further smooth fluctuations among individual data points, a moving average process can be applied before kernel density estimation, creating averages over a specified sampling window. In case where the data follows a degenerate distribution with all data points at a single value, we can model it using a Dirac delta function centred at that value, implemented as a very narrow Gaussian distribution with minimal variance.

% ------- Algorithm: Machine Unlearning -------
\begin{algorithm}[t]
\caption{Machine Unlearning}\label{alg:mac_unlearn}
\begin{algorithmic}
\State \textbf{Input}: unlearning set $\mathcal{D}\text{: }\{\mathcal{X}, \mathcal{Y}\}$ and hypothesis set $\mathcal{H}^*$
\State \textbf{Output}: updated model parameter $\boldsymbol{\theta}$
\State
\LineComment{\textit{initialisation}}
\State load model parameters $\theta$
\State set unlearning rate $\eta$

\State
\LineComment{\textit{machine unlearning via back-propagation}}
\While {convergence criterion is not satisfied}

\State generate pseudo-toxic samples $\tilde{\boldsymbol{x}} \gets (\boldsymbol{1} - \boldsymbol{m}^{*})\boldsymbol{x} + \boldsymbol{m}^{*}\boldsymbol{z}_y^{*}$ 
\State where $\boldsymbol{x} \in \mathcal{X}$ and $\{\boldsymbol{z}_y^{*}, \boldsymbol{m}\} \in \mathcal{H}^*$
\State compute gradient $\nabla_{\boldsymbol{\theta}} (\mathcal{L}(y, f_{\boldsymbol{\theta}}(\boldsymbol{x})) + \mathcal{L}(y, f_{\boldsymbol{\theta}}(\tilde{\boldsymbol{x}})))$
\State update $\boldsymbol{\theta} \gets \boldsymbol{\theta} - \eta \cdot \nabla_{\boldsymbol{\theta}} (\mathcal{L}(y, f_{\boldsymbol{\theta}}(\boldsymbol{x})) + \mathcal{L}(y, f_{\boldsymbol{\theta}}(\tilde{\boldsymbol{x}})))$
\EndWhile
\end{algorithmic}
\label{alg1}
\end{algorithm}

\subsection{Machine Unlearning}
Let $f_{\boldsymbol{\theta}}$ denote a potentially infected machine with its parameters $\boldsymbol{\theta}$ annotated explicitly. To unlearn the backdoor trigger while retaining the benign knowledge acquired previously, the machine is fine-tuned on both clean and pseudo-toxic samples from an unlearning set. The pseudo-toxic samples, denoted by $\widetilde{\mathcal{X}}$, are generated by
\begin{equation}
\tilde{\boldsymbol{x}} = (\boldsymbol{1} - \boldsymbol{m}^{*})\boldsymbol{x} + \boldsymbol{m}^{*}\boldsymbol{z}_y^{*} ,
\end{equation}
where the pair $\{\boldsymbol{z}_y^{*}$, $\boldsymbol{m}^{*}\}$ represents a selected hypothesis from $\mathcal{H}^*$ sampled with a probability inversely proportional to its average loss score. Note that the labels associated with the pseudo-toxic samples are assigned with the actual ground truth $y \in \mathcal{Y}$, instead of a hypothetical backdoor class. The machine's parameters are updated iteratively by
\begin{equation}
\boldsymbol{\theta}^{(t)} = \boldsymbol{\theta}^{(t-1)} - \eta \cdot \nabla_{\boldsymbol{\theta}} (\mathcal{L}(y, f_{\boldsymbol{\theta}}(\boldsymbol{x})) + \mathcal{L}(y, f_{\boldsymbol{\theta}}(\tilde{\boldsymbol{x}}))) .
\end{equation}
The machine unlearning process is outlined in Algorithm~\ref{alg:mac_unlearn}.

\section{Experiments}
We examine the proposed system in terms of fidelity, vulnerability and detectability on various datasets and neural network architectures with visual (qualitative) and numerical (quantitative) results. A comparative study is carried out to evaluate performance improvement upon the benchmarks.

\subsection{Experimental Setups}
For reproducibility and replicability, the experimental setups for the datasets, machine learning models and evaluation metrics are detailed as follows.

\subsubsection*{Datasets}
The experiments were conducted on two fundamental datasets for image classification in computer vision:
\begin{itemize}
	\item MNIST: This dataset consists of 70,000 grayscale images of 10 classes, each with a resolution of $28 \times 28$ pixels~\cite{726791}. The 10 classes represent handwritten digits from 0 to 9. We divide it into a learning set of 50,000 images, an inference set of 10,000 images and an auxiliary set of 10,000 images.
	\item CIFAR: This dataset consists of 60,000 colour images in 10 categories, each with a resolution of $32 \times 32$ pixels~\cite{krizhevsky2009learning}. The 10 classes represent aeroplane, automobile, bird, cat, deer, dog, frog, horse, ship, and truck. We divide it into a learning set of 40,000 images, an inference set of 10,000 images and an auxiliary set of 10,000 images.
	\item ImageNet: This dataset consists of over 1.2 million colour images across 1,000 categories with varying resolutions, which are typically resized to $224 \times 224$ pixels~\cite{5206848}. As a proof of concept, 10 classes are chosen to support rapid experimentation and benchmarking on a manageable subset of data, including the following objects: tench, English springer, cassette player, chain saw, church, French horn, garbage truck, gas pump, golf ball, and parachute. We divide it into a learning set of 9,469 images, an inference set of 3,825 images and an auxiliary set of 100 images.
\end{itemize}
The learning set was used for model training, whereas the inference set was used for deriving the experimental results, unless specified otherwise. The auxiliary set served as historical data for Bayesian statistics with a small portion of data used as the normative set for hypothesis analysis and as the unlearning set for machine unlearning. Both the number of samples used for hypothesis analysis and that for machine unlearning were fixed at $10$ samples per class (i.e. $100$ samples in total).

\begin{figure}[!t]
    \centering
    \subfloat{
        \includegraphics[width=0.45\textwidth]{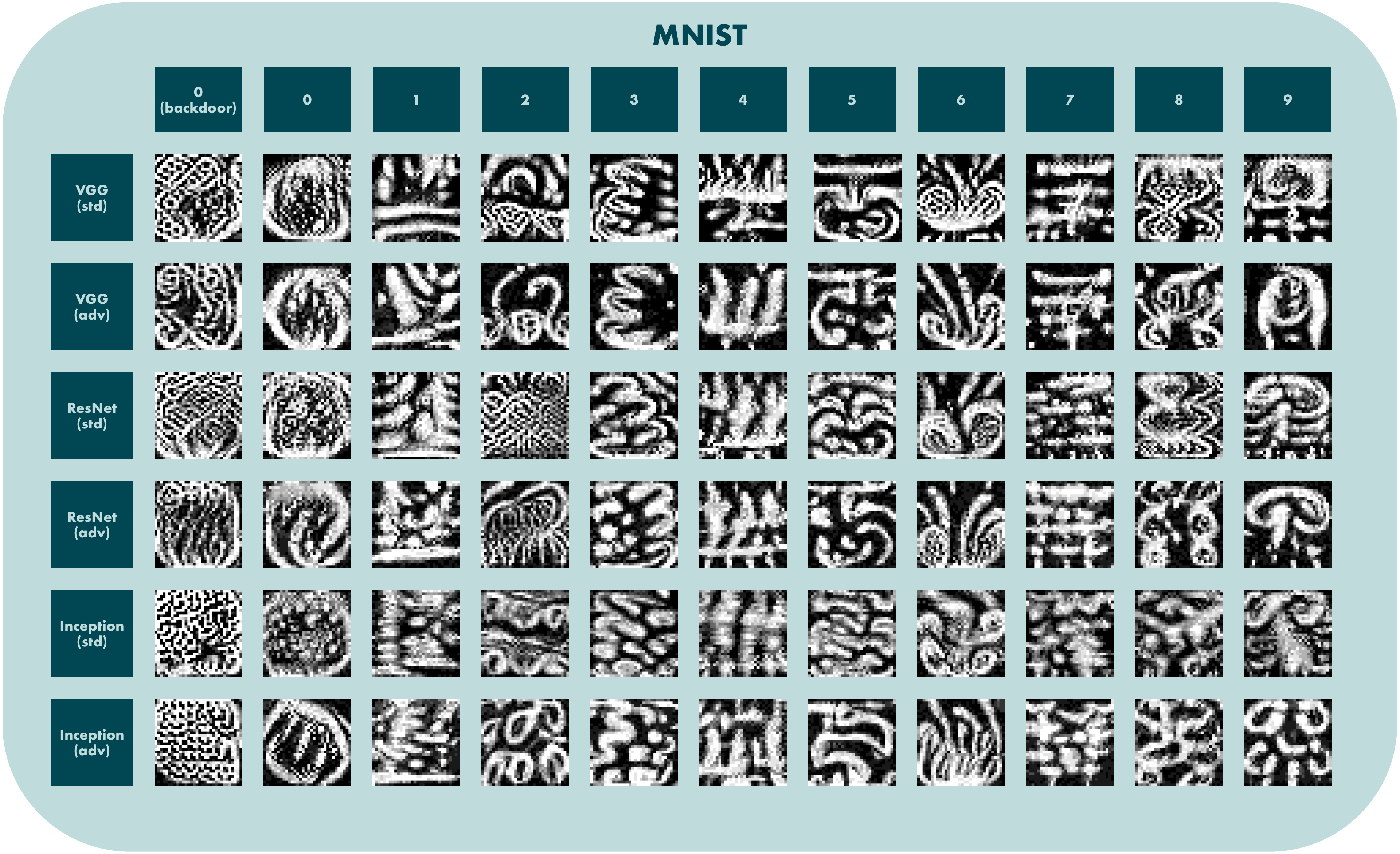}
        \label{fig:proto_mnist}
    }
    \\
    %\hspace{0.05\textwidth} % Space between the subfigures
    \subfloat{
        \includegraphics[width=0.45\textwidth]{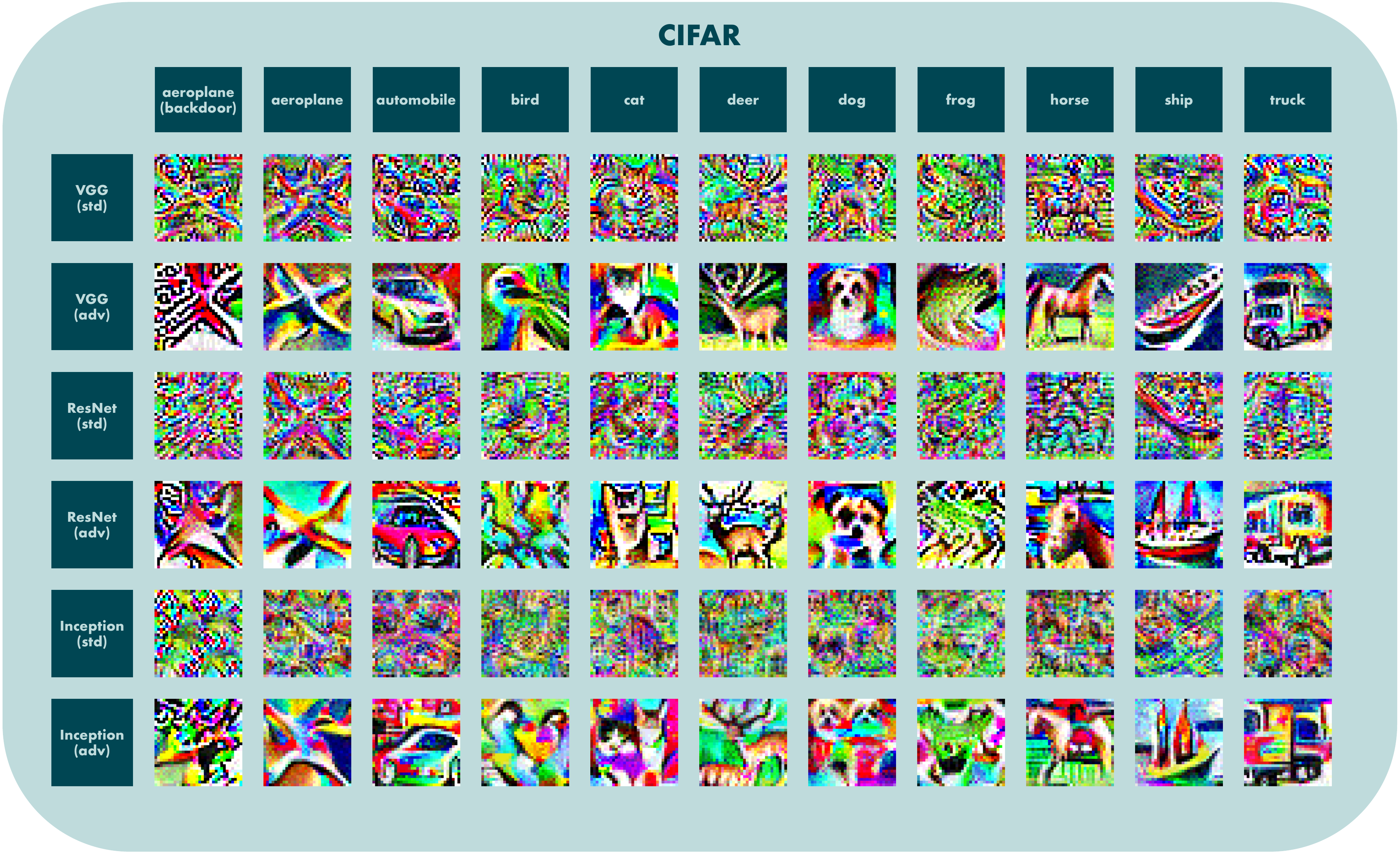}
        \label{fig:proto_cifar}
    }
    \\
    \subfloat{
        \includegraphics[width=0.45\textwidth]{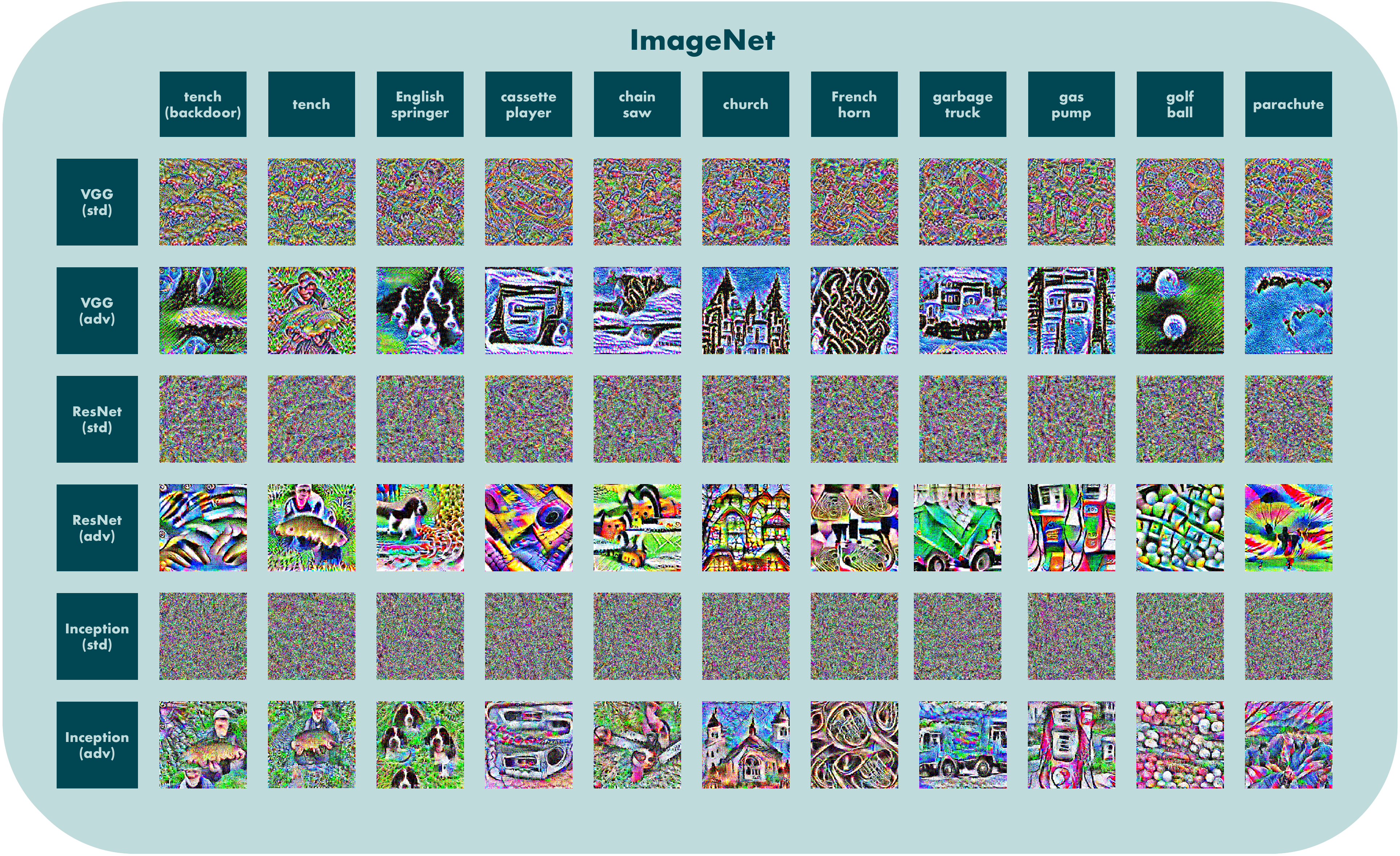}
        \label{fig:proto_imagenet}
    }
    \caption{Visualisation of artificial mental images from models of both infected and uninfected states.}
    \label{fig:proto}
\end{figure}

\subsubsection*{Models}
The selected models included three seminal convolutional neural network architectures:
\begin{itemize}
	\item VGG: This model emphasises simplicity and depth with small convolutional filters stacked throughout the entire neural network~\cite{7486599}.
	\item ResNet: This model contains residual connections, which act as shortcuts that bypass parameterised layers, allowing identity mappings for these layers~\cite{7780459}.
	\item Inception: This model combines multiple convolution paths with various kernel sizes and a max pooling operation in parallel, featuring a `network within a network' topology~\cite{7298594}.
\end{itemize}
For consistency, we unified the final part of each model as a concatenation of an adaptive average pooling layer and a fully connected layer. The former distills two-dimensional feature maps into a one-dimensional feature vector through summarising the spatial information into a single value. The latter acts as a classifier that applies linear combinations to map the feature vector into 10 logits, representing the unnormalised probabilities for 10 classes. Each model has two states, denoted as follows:
\begin{itemize}
	\item 0: An uninfected model trained on the benign samples.
	\item 1: An infected model trained on the malicious samples.
\end{itemize}
A backdoor attack was simulated with a poisoning rate of $5\%$ to ensure effective infection. Each model also involved two different learning paradigms, denoted as follows:
\begin{itemize}
	\item std: A model trained on the learning set within a standard learning paradigm.
	\item adv: A model trained on the learning set within an adversarial learning paradigm.
\end{itemize}

\begin{figure}[!t]
    \centering
    \subfloat{
        \includegraphics[width=0.42\textwidth]{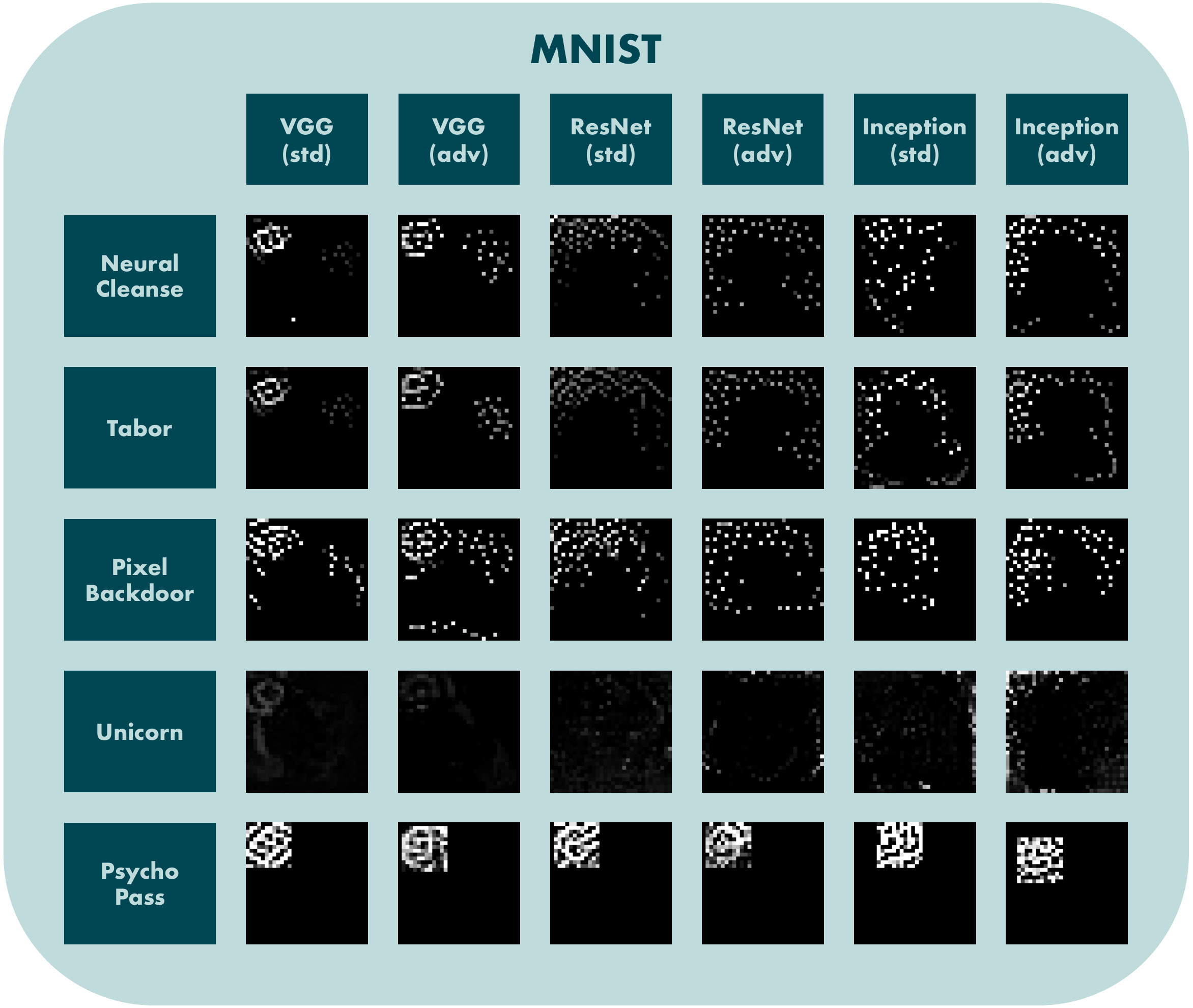}
        \label{fig:trigger_mnist}
    }
    \\
    %\hspace{0.05\textwidth} % Space between the subfigures
    \subfloat{
        \includegraphics[width=0.42\textwidth]{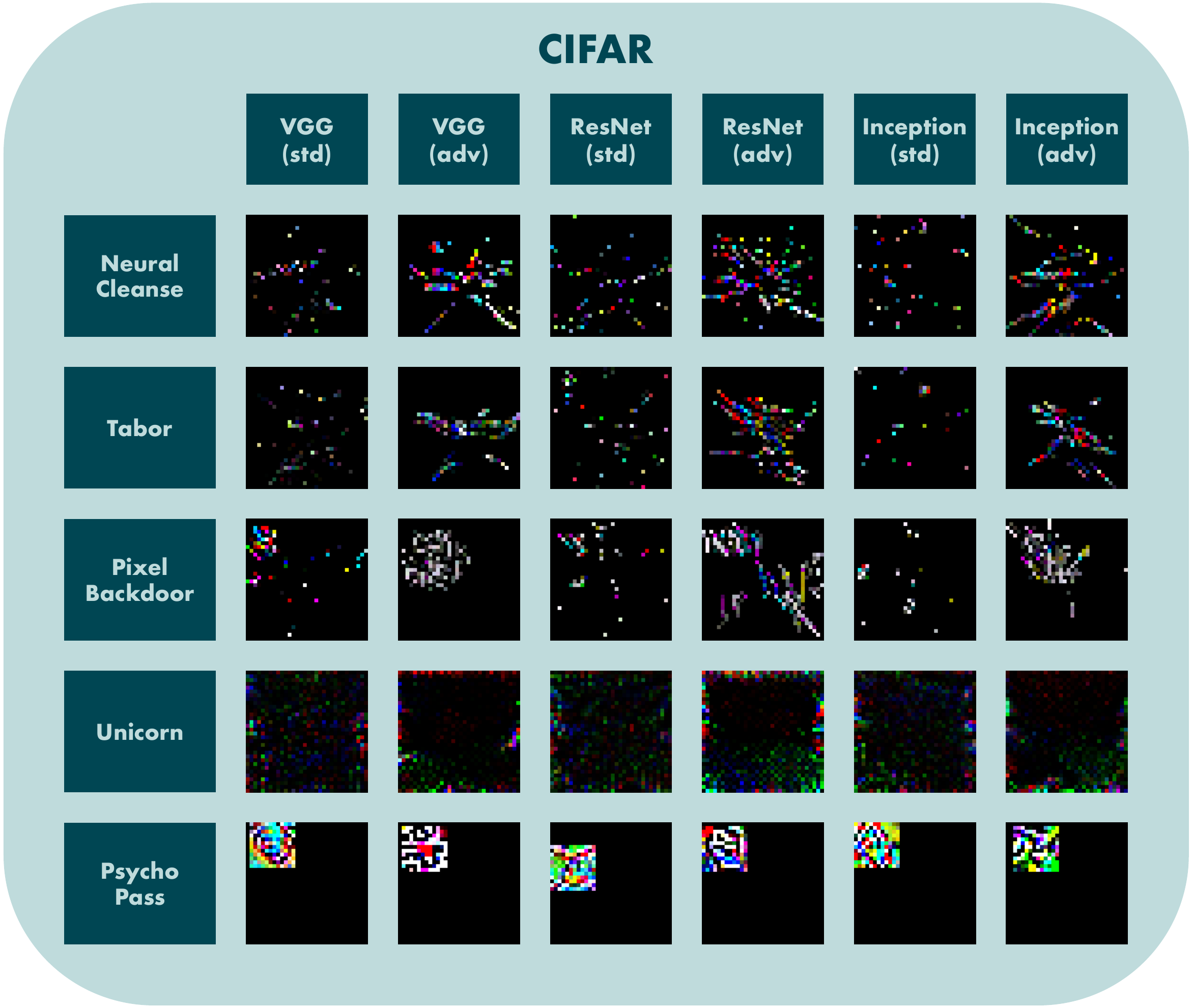}
        \label{fig:trigger_cifar}
    }
    \\
    \subfloat{
        \includegraphics[width=0.42\textwidth]{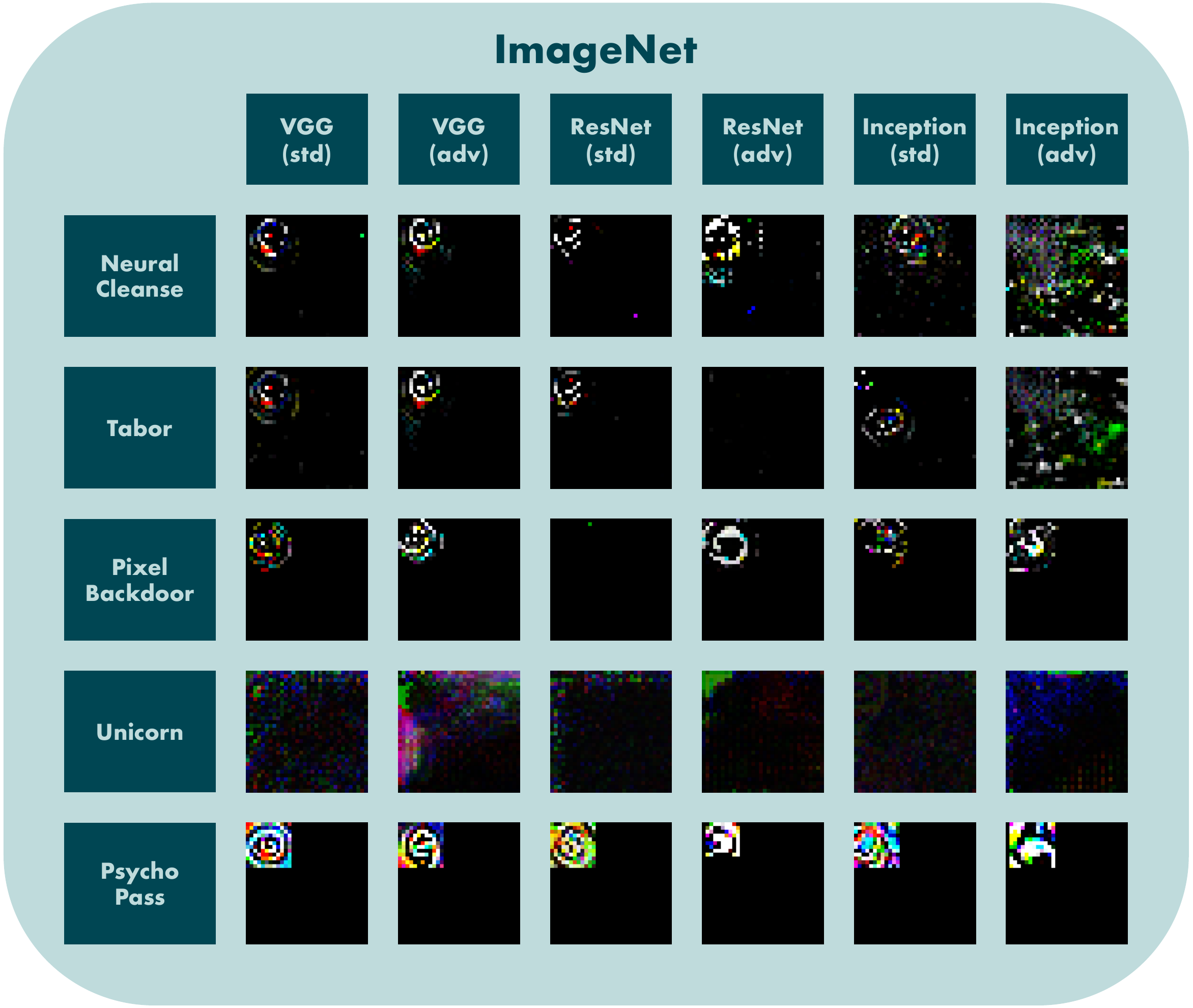}
        \label{fig:trigger_imagenet}
    }
    \caption{Visualisation of reverse-engineered triggers with various methods.}
    \label{fig:trigger}
\end{figure}

\subsubsection*{Metrics}
The primary evaluation metrics in this study were fidelity and vulnerability. Fidelity refers to the degree to which a processed model resembles the original model. We represent fidelity by comparing the classification accuracy (ACC) of the infected and disinfected models, against that of the uninfected models, defined as the number of correctly classified samples divided by the total number of samples:
\begin{equation}
	\text{ACC} = \frac{\text{correct classifications}}{\text{all classifications}} .
\end{equation}
Vulnerability refers to the extent to which a model can be manipulated or deceived into producing targeted predictions that align with an implanted backdoor. We measure vulnerability by the attack success rate (ASR), calculated as the number of toxic samples misclassified as the attack target class divided by the total number of samples excluding those inherently belonging to the attack target class:
\begin{equation}
	\text{ASR} = \frac{\text{misled classifications on toxic data}}{\text{all classifications except attack target}} .
\end{equation}

\subsubsection*{Benchmarks}
We selected 5 representative methods as benchmarks for our comparative study.
\begin{itemize}
	\item Fine-tuning: A na{\"i}ve baseline that directly updates the model parameters on a small amount of benign data, without any explicit backdoor mitigation strategy.
	\item Neural Cleanse: A method that optimises a pattern and a mask for each class, with the norm of the mask incorporated as a regularisation term~\cite{8835365}.
	\item Tabor: A method that optimises a pattern and a mask for each class, with various heuristic regularisation terms~\cite{9338311}.
	\item Pixel Backdoor: A method that optimises positive and negative perturbations, with regularisation on perturbation magnitude~\cite{9879000}.
	\item Unicorn: A method that jointly optimises a trigger pattern, a mask and a transformation function, with constraints on invertibility, stealthiness, mask size and activation disentanglement~\cite{wang2023unicorn}.
\end{itemize}

\subsubsection*{Hyperparameters:}
The following parameters were empirically defined. In model inversion, we set the number of artificial mental images per class as $20$, the step size of gradient descent as $0.1$, and the number of iterations as $50$ and the number of scales as $4$. In hypothesis analysis, we set the mask size as $12 \times 12$ pixels, the number of selected patterns as $20$, the radius for neighbouring patterns as $2$, the threshold for perceptual similarity as $0.1$, the threshold for the number of homogeneous cluster centroids as $1$ and the bandwidth for kernel density estimation as $0.5$. In machine unlearning, we set the number of epochs as $20$.

% Evaluations on MNIST
\begin{table*}[!t]
\caption{Evaluation of Fidelity \& Vulnerability on MNIST\label{tab:compare_mnist}}
\centering
%\begin{tabular}{lrr}
\begin{tabular}{l c c c c c c c c c c c c}
\toprule % VGG
Database & \multicolumn{12}{|c}{MNIST} \\
\midrule
Model & \multicolumn{4}{|c}{VGG} & \multicolumn{4}{|c}{ResNet} & \multicolumn{4}{|c}{Inception} \\
\midrule
Learning Mode & \multicolumn{2}{|c}{std} & \multicolumn{2}{|c}{adv} & \multicolumn{2}{|c}{std} & \multicolumn{2}{|c}{adv} & \multicolumn{2}{|c}{std} & \multicolumn{2}{|c}{adv} \\
\midrule
Metrics & \multicolumn{1}{|c}{\textbf{ACC} $\uparrow$} & \textbf{ASR} $\downarrow$ & \multicolumn{1}{|c}{\textbf{ACC} $\uparrow$} & \textbf{ASR} $\downarrow$ & \multicolumn{1}{|c}{\textbf{ACC} $\uparrow$} & \textbf{ASR} $\downarrow$ & \multicolumn{1}{|c}{\textbf{ACC} $\uparrow$} & \textbf{ASR} $\downarrow$ & \multicolumn{1}{|c}{\textbf{ACC} $\uparrow$} & \textbf{ASR} $\downarrow$ & \multicolumn{1}{|c}{\textbf{ACC} $\uparrow$} & \textbf{ASR} $\downarrow$ \\
\midrule
State 0 (uninfected) & \multicolumn{1}{|c}{0.9916} & 0.0000 & 0.9890 & 0.0000 & 0.9925 & 0.0003 & 0.9892 & 0.0002 & 0.9928 & 0.0004 & 0.9887 & 0.0003 \\
State 1 (infected) & \multicolumn{1}{|c}{0.9908} & 1.0000 & 0.9920 & 1.0000 & 0.9908 & 1.0000 & 0.9925 & 1.0000 & 0.9905 & 1.0000 & 0.9912 & 1.0000 \\
\midrule
Fine-Tuning & \multicolumn{1}{|c}{\cellcolor{green!20}0.9553} & 1.0000 & \cellcolor{green!20}0.9454 & 1.0000 & \cellcolor{green!20}0.9644 & 0.9836 & \cellcolor{green!20}0.9620 & 0.8970 & \cellcolor{green!20}0.9086 & 0.5846 & \cellcolor{green!20}\textbf{0.9812} & 1.0000\\
% \midrule
Neural Cleanse & \multicolumn{1}{|c}{\cellcolor{green!20}\textbf{0.9807}} & \cellcolor{green!20}0.0018 & \cellcolor{green!20}0.9679 & \cellcolor{green!20}0.0056 & \cellcolor{green!20}\textbf{0.9717} & 0.2530 & \cellcolor{green!20}0.9625 & 0.2602 & \cellcolor{green!20}\textbf{0.9751} & \cellcolor{green!20}\textbf{0.0003} & \cellcolor{green!20}0.9762 & 0.4594\\
Tabor & \multicolumn{1}{|c}{\cellcolor{green!20}0.9787} & \cellcolor{green!20}0.0012 & \cellcolor{green!20}0.9619 & \cellcolor{green!20}0.0544 & \cellcolor{green!20}0.9551 & 0.3299 & \cellcolor{green!20}0.9596 & 0.5820 & \cellcolor{green!20}0.9688 & \cellcolor{green!20}0.0044 & \cellcolor{green!20}0.9731 & 0.2307\\
Pixel Backdoor & \multicolumn{1}{|c}{\cellcolor{green!20}0.9697} & \cellcolor{green!20}0.0231 & \cellcolor{green!20}\textbf{0.9708} & \cellcolor{green!20}\textbf{0.0003} & \cellcolor{green!20}0.9592 & \cellcolor{green!20}0.0536 & \cellcolor{green!20}\textbf{0.9636} & 0.2519 & \cellcolor{green!20}0.9683 & \cellcolor{green!20}0.0224 & \cellcolor{green!20}0.9792 & \cellcolor{green!20}0.0283\\
Unicorn & \multicolumn{1}{|c}{\cellcolor{green!20}0.9553	} & \cellcolor{green!20}\textbf{0.0010} & \cellcolor{green!20}0.9315 & \cellcolor{green!20}0.0903 & \cellcolor{green!20}0.9464	 & 0.1549 & \cellcolor{green!20}0.9564 & \cellcolor{green!20}0.0067 & \cellcolor{green!20}0.9457 & \cellcolor{green!20}0.0065 & \cellcolor{green!20}0.9504 & \cellcolor{green!20}0.0108\\
Psycho-Pass (Ours) & \multicolumn{1}{|c}{\cellcolor{green!20}0.9778} & \cellcolor{green!20}0.0032 & \cellcolor{green!20}0.9665 & \cellcolor{green!20}0.0027 & \cellcolor{green!20}0.9639 & \cellcolor{green!20}\textbf{0.0293} & \cellcolor{green!20}0.9531 & \cellcolor{green!20}\textbf{0.0033} & \cellcolor{green!20}0.9630 & \cellcolor{green!20}0.0088 & \cellcolor{green!20}0.9732 & \cellcolor{green!20}\textbf{0.0012}\\
\bottomrule
\end{tabular}
\end{table*}

% Evaluations on CIFAR
\begin{table*}[!t]
\caption{Evaluation of Fidelity \& Vulnerability on CIFAR\label{tab:compare_cifar}}
\centering
%\begin{tabular}{lrr}
\begin{tabular}{l c c c c c c c c c c c c}
\toprule % VGG
Database & \multicolumn{12}{|c}{CIFAR} \\
\midrule
Model & \multicolumn{4}{|c}{VGG} & \multicolumn{4}{|c}{ResNet} & \multicolumn{4}{|c}{Inception} \\
\midrule
Learning Mode & \multicolumn{2}{|c}{std} & \multicolumn{2}{|c}{adv} & \multicolumn{2}{|c}{std} & \multicolumn{2}{|c}{adv} & \multicolumn{2}{|c}{std} & \multicolumn{2}{|c}{adv} \\
\midrule
Metrics & \multicolumn{1}{|c}{\textbf{ACC} $\uparrow$} & \textbf{ASR} $\downarrow$ & \multicolumn{1}{|c}{\textbf{ACC} $\uparrow$} & \textbf{ASR} $\downarrow$ & \multicolumn{1}{|c}{\textbf{ACC} $\uparrow$} & \textbf{ASR} $\downarrow$ & \multicolumn{1}{|c}{\textbf{ACC} $\uparrow$} & \textbf{ASR} $\downarrow$ & \multicolumn{1}{|c}{\textbf{ACC} $\uparrow$} & \textbf{ASR} $\downarrow$ & \multicolumn{1}{|c}{\textbf{ACC} $\uparrow$} & \textbf{ASR} $\downarrow$ \\
\midrule
State 0 (uninfected) & \multicolumn{1}{|c}{0.9085} & 0.0096 & 0.7884 & 0.0104 & 0.9171 & 0.0194 & 0.8309 & 0.0162 & 0.9177 & 0.0175 & 0.8360 & 0.0021\\
State 1 (infected) & \multicolumn{1}{|c}{0.9028} & 0.9993 & 0.7930 & 1.0000 & 0.9126 & 0.9996 & 0.8163 & 1.0000 & 0.9091 & 0.9993 & 0.8328 & 1.0000 \\
\midrule
Fine-Tuning & \multicolumn{1}{|c}{\cellcolor{green!20}\textbf{0.8465}} & 1.0000 & \textbf{0.6472} & 0.9991 & 0.7856 & 0.7368 & \textbf{0.6422} & 0.9782 & \textbf{0.7950} & 1.0000 & \cellcolor{green!20}\textbf{0.7513} & 1.0000 \\
% \midrule
Neural Cleanse & \multicolumn{1}{|c}{\cellcolor{green!20}0.8337} & 0.2443 & 0.6044 & 0.8328 & 0.7427 & 0.7448 & 0.6158 & 0.5937 & 0.7604 & 0.9599 & 0.7231 & 1.0000\\
Tabor & \multicolumn{1}{|c}{\cellcolor{green!20}0.8458} & 0.8507 & 0.6163 & 0.9145 & \textbf{0.7981} & 0.8069 & 0.6204 & 0.9595 & 0.7938 & 0.9962 & 0.7316 & 1.0000\\
Pixel Backdoor & \multicolumn{1}{|c}{\cellcolor{green!20}0.8332} & 0.0549 & 0.5899 & 0.0919 & 0.7728 & 0.3359 & 0.5809 & 0.5990 & 0.7761 & 0.8831 & 0.6891 & \cellcolor{green!20}0.0910 \\
Unicorn & \multicolumn{1}{|c}{0.7989	} & 0.8151 & 0.5715 & 0.1002 & 0.7343 & 0.5040 & 0.5797 & 0.9933 & 0.7646 & 0.5066 & 0.6905 & 1.000\\
Psycho-Pass (Ours) & \multicolumn{1}{|c}{\cellcolor{green!20}0.8389} & \cellcolor{green!20}\textbf{0.0121} & 0.6304 & \cellcolor{green!20}\textbf{0.0353} & 0.7894 & \textbf{0.2037} & 0.6302 & \cellcolor{green!20}\textbf{0.0924} & 0.7615 & \cellcolor{green!20}\textbf{0.0709} & \cellcolor{green!20}0.7395 & \cellcolor{green!20}\textbf{0.0514}\\
\bottomrule
\end{tabular}
\end{table*}

% Evaluations on ImageNet
\begin{table*}[!t]
\caption{Evaluation of Fidelity \& Vulnerability on ImageNet\label{tab:compare_imagenet}}
\centering
%\begin{tabular}{lrr}
\begin{tabular}{l c c c c c c c c c c c c}
\toprule % VGG
Database & \multicolumn{12}{|c}{ImageNet} \\
\midrule
Model & \multicolumn{4}{|c}{VGG} & \multicolumn{4}{|c}{ResNet} & \multicolumn{4}{|c}{Inception} \\
\midrule
Learning Mode & \multicolumn{2}{|c}{std} & \multicolumn{2}{|c}{adv} & \multicolumn{2}{|c}{std} & \multicolumn{2}{|c}{adv} & \multicolumn{2}{|c}{std} & \multicolumn{2}{|c}{adv} \\
\midrule
Metrics & \multicolumn{1}{|c}{\textbf{ACC} $\uparrow$} & \textbf{ASR} $\downarrow$ & \multicolumn{1}{|c}{\textbf{ACC} $\uparrow$} & \textbf{ASR} $\downarrow$ & \multicolumn{1}{|c}{\textbf{ACC} $\uparrow$} & \textbf{ASR} $\downarrow$ & \multicolumn{1}{|c}{\textbf{ACC} $\uparrow$} & \textbf{ASR} $\downarrow$ & \multicolumn{1}{|c}{\textbf{ACC} $\uparrow$} & \textbf{ASR} $\downarrow$ & \multicolumn{1}{|c}{\textbf{ACC} $\uparrow$} & \textbf{ASR} $\downarrow$ \\
\midrule
State 0 (uninfected) & \multicolumn{1}{|c}{0.9842} & 0.0000 & 0.8495 & 0.0109 & 0.9871 & 0.0006 & 0.8797 & 0.0153 & 0.9813 & 0.0006 & 0.8744 & 0.0391\\
State 1 (infected) & \multicolumn{1}{|c}{0.9803} & 1.0000 & 0.6539 & 1.0000 & 0.9834 & 0.9997 & 0.8799 & 0.9994 & 0.9811 & 1.0000 & 0.8705 & 1.0000 \\
\midrule
Fine-Tuning & \multicolumn{1}{|c}{\cellcolor{green!20}0.9139} & 0.9941 & \cellcolor{green!20}0.5574 & 0.9947 & \cellcolor{green!20}0.9400 & 0.8903 & 0.7639 & 0.9993 & \cellcolor{green!20}0.9713 & 1.0000 & \cellcolor{green!20}0.8816 & 1.0000\\
% \midrule
Neural Cleanse & \multicolumn{1}{|c}{\cellcolor{green!20}0.9113} & \cellcolor{green!20}0.0000 & 0.5161 & \cellcolor{green!20}0.0326 & \cellcolor{green!20}0.9334 & \cellcolor{green!20}0.0076 & \cellcolor{green!20}0.7976 & \cellcolor{green!20}0.0432 & \cellcolor{green!20}0.9524 & 0.2079 & \cellcolor{green!20}0.8579 & 1.0000\\
Tabor & \multicolumn{1}{|c}{\cellcolor{green!20}0.9118} & \cellcolor{green!20}0.0005 & \cellcolor{green!20}\textbf{0.5752} & \cellcolor{green!20}0.0934 & \cellcolor{green!20}0.9423 & \cellcolor{green!20}0.0235 & \cellcolor{green!20}0.8094 & 0.9997 & \cellcolor{green!20}0.9597 & 0.3010 & \cellcolor{green!20}\textbf{0.8903} & 1.0000\\
Pixel Backdoor & \multicolumn{1}{|c}{\cellcolor{green!20}0.8853} & \cellcolor{green!20}0.0056 & \cellcolor{green!20}0.5672 & \cellcolor{green!20}\textbf{0.0406} & \cellcolor{green!20}\textbf{0.9611} & \cellcolor{green!20}0.0406 & \cellcolor{green!20}0.8037 & \cellcolor{green!20}0.0962 & \cellcolor{green!20}\textbf{0.9671} & 0.1932 & \cellcolor{green!20}0.8745 & 0.3379 \\
Unicorn & \multicolumn{1}{|c}{0.8099} & \cellcolor{green!20}0.0231 & 0.5323 & 0.1320 & \cellcolor{green!20}0.9346 & 0.3262 & 0.7589 & \cellcolor{green!20}0.9997 & \cellcolor{green!20}0.9519 & 0.8805 & \cellcolor{green!20}0.8335 & 0.9985\\
Psycho-Pass (Ours) & \multicolumn{1}{|c}{\cellcolor{green!20}\textbf{0.9158}} & \cellcolor{green!20}\textbf{0.0097} & 0.5379 & \cellcolor{green!20}0.0626 & \cellcolor{green!20}0.9339 & \cellcolor{green!20}\textbf{0.0021} & \cellcolor{green!20}\textbf{0.8168} & \cellcolor{green!20}\textbf{0.0344} & \cellcolor{green!20}\textbf{0.9671} & \cellcolor{green!20}\textbf{0.0594} & \cellcolor{green!20}0.8871 & \textbf{0.1664}\\
\bottomrule
\end{tabular}
\end{table*}

% Bayesian Inference on MNIST
\begin{table*}[ht]
\caption{Evaluation of Detectability on MNIST\label{tab:bayes_mnist}}
\centering
%\begin{tabular}{lrr}
\begin{tabular}{l c c c c c c c c c c c c}
\toprule % VGG
Database & \multicolumn{12}{|c}{MNIST} \\
\midrule
Model & \multicolumn{4}{|c}{VGG} & \multicolumn{4}{|c}{ResNet} & \multicolumn{4}{|c}{Inception} \\
\midrule
Learning Mode & \multicolumn{2}{|c}{std} & \multicolumn{2}{|c}{adv} & \multicolumn{2}{|c}{std} & \multicolumn{2}{|c}{adv} & \multicolumn{2}{|c}{std} & \multicolumn{2}{|c}{adv} \\
\midrule
State & \multicolumn{1}{|c}{0} & 1 & \multicolumn{1}{|c}{0} & 1 & \multicolumn{1}{|c}{0} & 1 & \multicolumn{1}{|c}{0} & 1 & \multicolumn{1}{|c}{0} & 1 & \multicolumn{1}{|c}{0} & 1 \\
\midrule
Probability of Infection & \multicolumn{1}{|c}{0.0036} & 1.0000 & 0.0024 & 1.0000 & 0.1242 & 1.0000 & 0.0000 & 1.0000 & 0.0002 & 1.0000 & 0.0001 & 1.0000\\
\bottomrule
\end{tabular}
\end{table*}

% Bayesian Inference on CIFAR
\begin{table*}[ht]
\caption{Evaluation of Detectability on CIFAR\label{tab:bayes_cifar}}
\centering
%\begin{tabular}{lrr}
\begin{tabular}{l c c c c c c c c c c c c}
\toprule % VGG
Database & \multicolumn{12}{|c}{CIFAR} \\
\midrule
Model & \multicolumn{4}{|c}{VGG} & \multicolumn{4}{|c}{ResNet} & \multicolumn{4}{|c}{Inception} \\
\midrule
Learning Mode & \multicolumn{2}{|c}{std} & \multicolumn{2}{|c}{adv} & \multicolumn{2}{|c}{std} & \multicolumn{2}{|c}{adv} & \multicolumn{2}{|c}{std} & \multicolumn{2}{|c}{adv} \\
\midrule
State & \multicolumn{1}{|c}{0} & 1 & \multicolumn{1}{|c}{0} & 1 & \multicolumn{1}{|c}{0} & 1 & \multicolumn{1}{|c}{0} & 1 & \multicolumn{1}{|c}{0} & 1 & \multicolumn{1}{|c}{0} & 1 \\
\midrule
Probability of Infection & 0.0115 & 0.8620 & 0.2417 & 1.0000 & 0.2052 & 1.0000 & \makebox[\widthof{\textbf{0.9999}}][c]{n/a} & 0.9959 & 0.0001 & 1.0000 & \makebox[\widthof{\textbf{0.9999}}][c]{n/a} & 1.0000 \\
\bottomrule
\end{tabular}
\end{table*}

% and a baseline (simple fine-tuning on benign samples) 

\subsection{Artificial Mental Imagery}
The artificial mental images projected from each model are visualised in Figure~\ref{fig:proto}. Specifically, the projections of the backdoor class distinctly exhibited a blend of features from both the benign samples and the backdoor trigger. Furthermore, models trained under adversarial conditions produced images that were less noisy and more visually interpretable, highlighting the backdoor features more clearly. This suggests that while adversarial training may lead to a reduction in classification accuracy, it can potentially provide a more precise characterisation of backdoor triggers for forensic analysis.

% This capability is crucial in ensuring that the unlearning process is targeted and effective in mitigating the impact of the malicious triggers.

\subsection{Fidelity $\&$ Vulnerability}
The experiments were conducted to evaluate the effectiveness of our proposed method in reverse engineering and unlearning backdoor triggers. We compared our approach with four benchmark methods and a baseline (simple fine-tuning on benign samples) across multiple scenarios involving three datasets (MNIST, CIFAR and ImageNet), three model architectures (VGG, ResNet and Inception) and two training modes (standard and adversarial). We visualised the reverse-engineered triggers generated by our method and the four benchmarks in Figure~\ref{fig:trigger}, where the triggers for ImageNet are magnified by cropping the top-left $32 \times 32$ region from the original $224 \times 224$ images to enhance visibility. The visualisations reveal that our method not only localises the position of the triggers more accurately but also recovers a pattern that is visually closer to the actual trigger. This precise reconstruction is essential for ensuring the reliability and specificity of the unlearning process. The benchmarks, while occasionally able to reverse engineer the trigger, often produced less precise and less similar patterns. In addition to this, we measured fidelity through ACC and vulnerability through ASR across three datasets, as shown in Table~\ref{tab:compare_mnist}, Table~\ref{tab:compare_cifar} and Table~\ref{tab:compare_imagenet}, respectively. The ACCs are marked in green if they drop by no more than 10\% relative to the accuracy of the unprocessed infected model, indicating sufficiently high fidelity preservation. The ASRs are marked in green if their values are below 10\%, indicating sufficiently low backdoor vulnerability. Across all tested configurations, the fidelity scores were comparable between our method and the benchmarks. This consistency is expected since all methods involved fine-tuning the models for the same number of epochs using the same set of benign samples. In terms of the vulnerability scores, our method consistently outperformed the benchmarks. The vulnerability scores for our method were consistently low, indicating successful backdoor removal. In contrast to this, the benchmark methods occasionally failed to eliminate the backdoors, resulting in higher vulnerability scores. The baseline method, relying solely on simple fine-tuning, was proved ineffective in unlearning the triggers, as the vulnerability scores remained high.

\begin{figure}[t]
    \centering
    \subfloat{
        \includegraphics[width=0.4\textwidth]{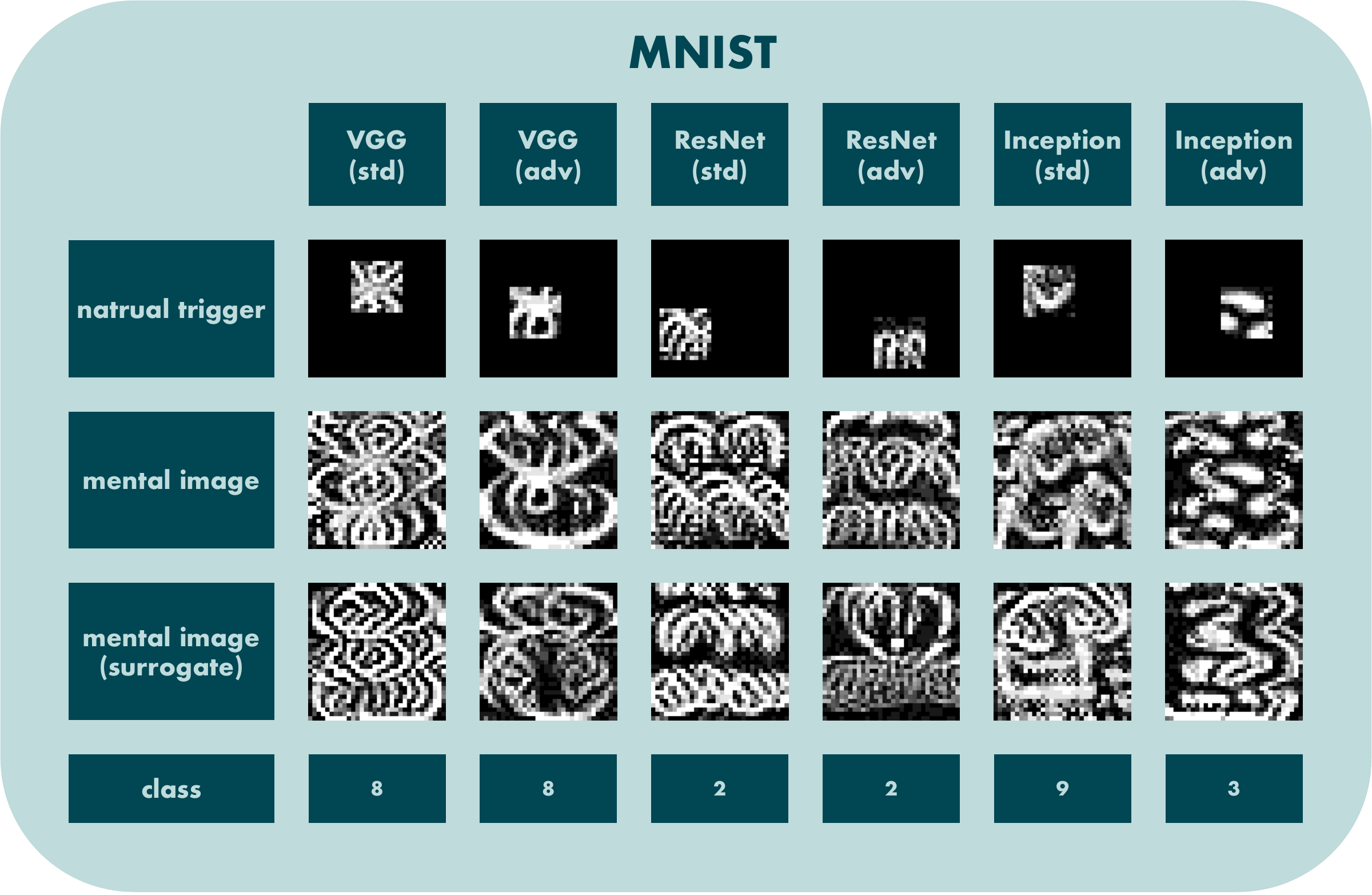}
        \label{fig:nat_trigger_mnist}
    }
    \\
    %\hspace{0.05\textwidth} % Space between the subfigures
    \subfloat{
        \includegraphics[width=0.4\textwidth]{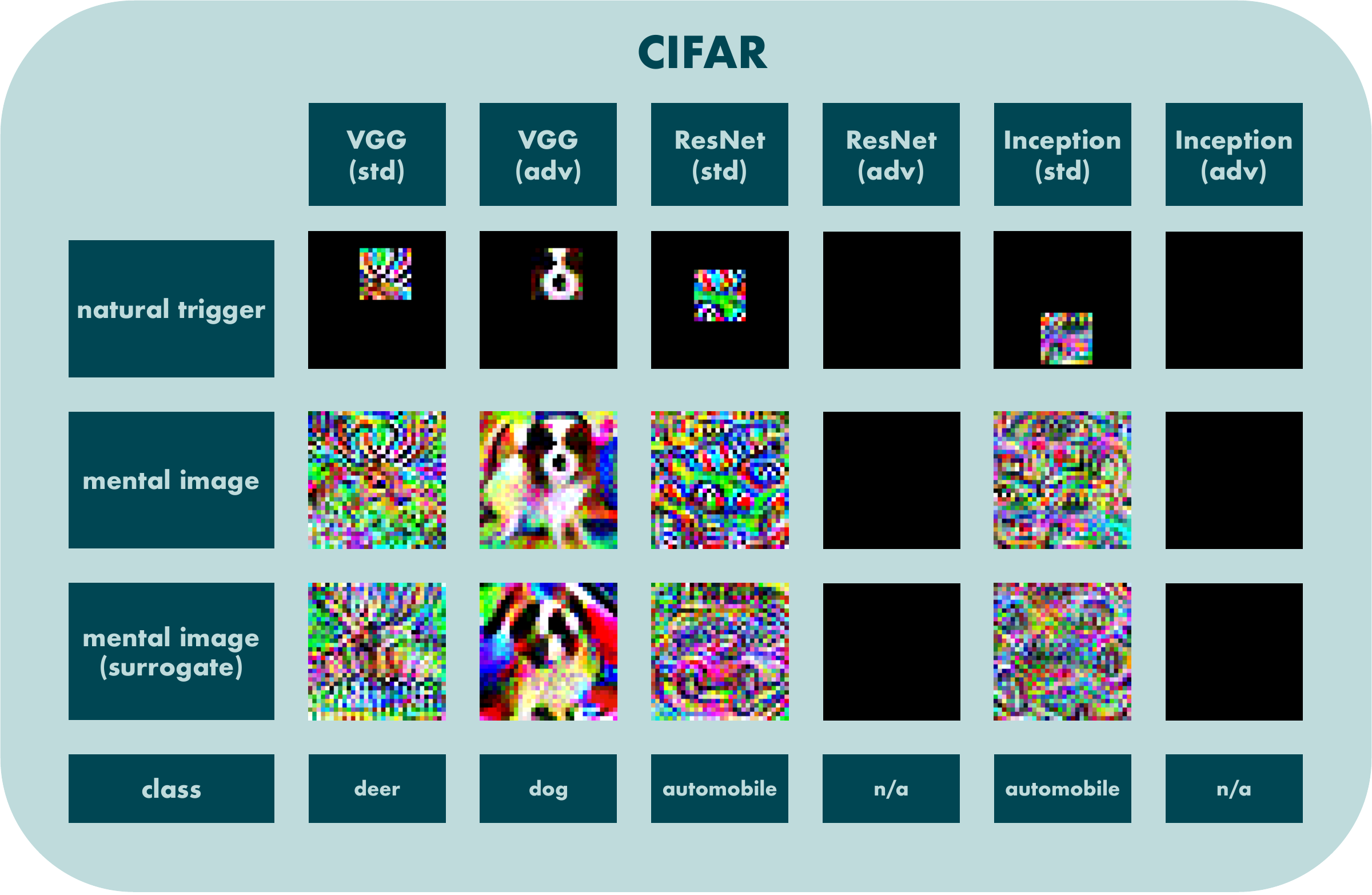}
        \label{fig:nat_trigger_cifar}
    }
    \caption{Visualisation of natural triggers along with artificial mental images from test and surrogate models of uninfected state.}
    \label{fig:nat_trigger}
\end{figure}

\subsection{Detectability}
We evaluated the backdoor detectability of Bayesian inference by analysing the estimated probabilities of infection for both infected and uninfected models, as detailed in Tables~\ref{tab:bayes_mnist} and~\ref{tab:bayes_cifar}. Cases where no triggers were detected following the outlier exclusion process are marked as not applicable (n/a), indicating a tendency to favour the non-infection decision. In contrast to MNIST and CIFAR, no outliers were observed in the ImageNet experiments, likely because the spatial resolution of ImageNet images is substantially larger than that of the hypothetical trigger, making it difficult for a small spurious pattern to mislead classification outcomes. The results indicate that backdoors were effectively detected even in the presence of persistent outliers that were not removed during the outlier exclusion process. Furthermore, the most likely hypotheses from uninfected models, including the selected triggers and their corresponding artificial mental images, are visualised in Figure~\ref{fig:nat_trigger}. Each trigger reflected prototypical patterns for a particular class and is therefore considered a natural trigger. The artificial mental images retrieved from uninfected models resembled those from their surrogate counterparts, confirming the validity of probabilistic inference through perceptual analysis.

\subsection{Limitations}
Our method specifically targets a typical type of backdoor trigger, enabling us to achieve reliable results for this common attack scenario, which represents a significant portion of real-world cases. Nevertheless, it is important to recognise the diversity of trigger patterns observed in the wild. Additionally, we assume that prior knowledge about trigger dimensions is available, which may not always be the case in practice. These limitations highlight areas for future work, such as extending the method to accommodate unknown or varying trigger dimensions and broadening the scope to cover a wider variety of backdoor patterns.

\section{Conclusion}
In this study, we investigated a cybernetic framework for automated surveillance of backdoor threats, recognising the dynamic nature of data sources. We proposed a methodology for detecting and unlearning backdoors implanted into neural network machines. In particular, we employed model inversion to project artificial mental images of each possible response from a machine, and conducted hypothesis analysis to infer the likelihood of each hypothetically malicious pattern being the true backdoor trigger. Based upon the feedback from statistical inference, the machine unlearning process is autonomously activated to dissociate the machine's behaviour from the estimated backdoor trigger. Experimental results demonstrate a stable balance between knowledge fidelity and backdoor vulnerability. The detectability evaluation validates the effectiveness of probabilistic inference through perceptual analysis of artificial mental images. Future research is essential to reliably address in-the-wild attack scenarios where trigger dimensions and patterns may be varied and elusive. Furthermore, it is crucial to investigate the characteristics of extrinsic backdoor triggers and intrinsic natural triggers, and to propose robust solutions for effectively separating one from the other.

\section*{Acknowledgement}
This work was supported in part by the Japan Society for the Promotion of Science (JSPS) under KAKENHI Grants (JP21H04907 and JP24H00732), and in part by the Japan Science and Technology Agency (JST) under CREST Grant (JPMJCR20D3) including AIP Challenge Program, AIP Acceleration Grant (JPMJCR24U3) and K Program Grant (JPMJKP24C2).

 %\newpage
\bibliography{Transactions-Bibliography/bstcontrol, Bib/bib_backdoor}
\bibliographystyle{Transactions-Bibliography/IEEEtran}

%\vfill

\vspace{4em}
\begin{IEEEbiographynophoto}
{Ching-Chun Chang} 
received the PhD in Computer Science from the University of Warwick, UK, in 2019. He is currently affiliated with the National Institute of Informatics, Japan, as a Project Assistant Professor. He participated in the Short-Term Scientific Mission supported by European Cooperation in Science and Technology Actions at the Faculty of Computer Science, Otto von Guericke University of Magdeburg, Germany, in 2016. He was granted the Marie-Curie Fellowship and participated in the Research and Innovation Staff Exchange supported by Marie Skłodowska-Curie Actions at the Department of Electrical and Computer Engineering, New Jersey Institute of Technology, USA, in 2017. He was a Visiting Scholar at the School of Computing and Mathematics, Charles Sturt University, Australia, in 2018, and at the School of Information Technology, Deakin University, Australia, in 2019. He was a Research Fellow at the Department of Electronic Engineering, Tsinghua University, China, in 2020. His research interests include artificial intelligence, biometrics, cryptography, cybersecurity, evolutionary computation, forensics, information theory, steganography, and watermarking.
\end{IEEEbiographynophoto}

\newpage
\begin{IEEEbiographynophoto}
{Kai Gao} received his BSc degree in Software Engineering from Fujian Normal University, China, in 2018. He is currently pursuing the PhD degree in the Department of Information Engineering and Computer Science, Feng Chia University. His research interests include cybersecurity, cryptography, secret sharing, steganography and machine learning.
\end{IEEEbiographynophoto}

\vspace{-20pt}
\begin{IEEEbiographynophoto}
{Shuying Xu} received her BSc degree in Software Engineering from Fujian Normal University, China, in 2019. She is currently pursuing the PhD degree in the Department of Information Engineering and Computer Science, Feng Chia University. Her research interests include cybersecurity, cryptography, biometrics, steganography and machine learning.
\end{IEEEbiographynophoto}

\vspace{-20pt}
\begin{IEEEbiographynophoto}
{Anastasia Kordoni} received her PhD degree in Psychology from Lancaster University, UK, in 2021. She is currently a Senior Research Associate with Fylde College, Lancaster University, UK. She is interested in the psychosocial underpinnings of resilience in security and safety settings. Her research focuses on understanding how people avoid, adapt to, and recover from external disruptions and adversity, and the role of emerging technologies in shaping resilient responses. She is working on an interdisciplinary project funded by the EPSRC Trustworthy Autonomous Systems Hub on the integration of social psychological principles into robotic systems for better emergency response and a synthesis of a Decision-from-Evidence framework to guide future questions and opportunities for emergency resilience. Her research has implications for policy and practice within different domains and contributes to theoretical advances of group behaviour.
\end{IEEEbiographynophoto}

\vspace{-20pt}
\begin{IEEEbiographynophoto}
{Christopher Leckie} is a Professor with the School of Computing and Information Systems, The University of Melbourne. His research interests include artificial intelligence (AI), machine learning, anomaly detection, unsupervised learning, telecommunications, and cybersecurity. He has a strong interest in developing AI and machine learning techniques for a variety of applications in telecommunications, such as cyber security, network management, fault diagnosis, and the Internet-of-Things. He also has an interest in robust and scalable machine learning algorithms for problems such as clustering and anomaly detection, with a focus on adversarial machine learning and security analytics.
\end{IEEEbiographynophoto}

\vspace{-20pt}
\begin{IEEEbiographynophoto}
{Isao Echizen}
received BS, MS, and DE degrees from the Tokyo Institute of Technology, Japan, in 1995, 1997 and 2003, respectively. He joined Hitachi, Ltd. in 1997 and until 2007 was a Research Engineer in the company's systems development laboratory. He is currently a Director and Professor of the Information and Society Research Division, as well as a Director of the Global Research Center for Synthetic Media, at the National Institute of Informatics; a Professor in the Department of Information and Communication Engineering, Graduate School of Information Science and Technology, the University of Tokyo; and a Professor in the Graduate University for Advanced Studies (SOKENDAI), Japan. He was a Visiting Professor at the Tsuda University, Japan; at the University of Freiburg, Germany; and at the University of Halle-Wittenberg, Germany. He is currently engaged in research on AI security, multimedia security and multimedia forensics, serving as a Research Director for the CREST FakeMedia project and the K Program SYNTHETIQ X project of the Japan Science and Technology Agency (JST). He received the Commendation for Science and Technology by the Minister of Education, Culture, Sports, Science and Technology (Research Category) in 2025. He also received the IEICE Best Paper Award in 2023; the IPSJ Best Paper Awards in 2005 and 2014; the IPSJ Nagao Special Researcher Award in 2011; the DOCOMO Mobile Science Award in 2014; the IISEC Information Security Cultural Award in 2016; and the IEEE WIFS Best Paper Award in 2017. He is an IEICE Fellow, an IPSJ Fellow, an IEEE Senior Member, an IFIP Japanese Representative, and an APSIPA Vice President.
\end{IEEEbiographynophoto}

\end{document}